# A scaling relation in [C II]-detected galaxies and its likely application in cosmology

Yi-Han Wu,★ Yu Gao† and Jun-Feng Wang★

*Department of Astronomy, College of Physical Science and Technology, Xiamen University, Xiamen 361005, China*

**ABSTRACT**
We identify and investigate a possible correlation between the [C II] 158−μm luminosity and linewidth in the [C II]-detected galaxies. Observationally, the strength of the [C II] 158−μm emission line is usually stronger than that of the carbon monoxide (CO) emission line and this [C II] line has been used as another tracer of the galactic characteristics. Moreover, many [C II]-detected galaxies are identified in $z > 4$. Motivated by previous studies of the CO luminosity–full width at half-maximum correlation relation (LFR) and the available new [C II] measurements, we compile samples of the [C II]-detected galaxies in the literature and perform the linear regression analysis. The [C II] LFR is confirmed at a robust level. We also demonstrate the possible application of the [C II] LFR by utilizing it on the distance measurement of the high-$z$ galaxy. As a result, we extend the cosmic spatial scale beyond the redshift $z$ of 7. With the outcome of the distance measurement, we constrain the cosmology parameters in the Chevallier–Polarski–Linder model, which considers the evolution of dark energy. Consequently, the uncertainties of the $w_0$ and $w_a$ are reduced significantly when the measured distance data of the [C II]-detected galaxies are included in the cosmological parameter constraint, exemplifying the potential of using the [C II]-detected galaxies as a tracer to constrain the cosmological parameters.

## 1 INTRODUCTION

The infrared (IR) and submillimeter survey projects implemented by the current high-technical observatories and arrays (e.g. *ALMA*, *Herschel*, *NOEMA*, and *SOFIA*) have dramatically improved the galaxy numbers observed at different redshifts of our Universe. By detecting the atomic and molecular lines emitting from the galaxies, one allows measuring the characteristics (such as molecular gas mass, luminosity, and kinematic velocity) of the galaxies in different cosmic redshifts (e.g. Bouwens et al. 2011; Geach et al. 2011; Sargent et al. 2012). For instance, the emission lines generated from the transitions of the rotational state $J$ ($J + 1 \rightarrow J$, and $J = 1, 2, 3$, and so on) of carbon monoxide (CO) are usually detected in the galaxy observations, and have been used as a tracer of galaxy characteristics in different cosmic redshifts (e.g. Blain 2002; Bolatto, Wolfire & Leroy 2013; Carilli & Walter 2013; Casey, Narayanan & Cooray 2014; Hodge & da Cunha 2020).

Some previous works discussed the existence of a correlation between the CO-line luminosity and full width at half-maximum (FWHM) linewidth in the CO-detected galaxies [i.e. a CO luminosity–FWHM linewidth correlation relation, and we denotes this relation as CO LFR hereafter; e.g. Harris et al. 2012; Bothwell et al. 2013; Goto & Toft 2015 (hereafter G15); Aravena et al. 2016; Sharon et al. 2016; Tiley et al. 2016; Isbell, Xue & Fu 2018; Topal et al. 2018; Neri et al. 2020].

Among the studies about the CO LFR, the CO luminosity ($L'_{\rm CO}$) is utilized to measure the molecular gas mass ($M_{\rm mol}$) of a galaxy, and the CO FWHM linewidth (FWHM$_{\rm CO}$, in km s$^{-1}$, which is sensitive to the self-structure dynamical velocity of a galaxy) is used to measure the dynamical mass ($M_{\rm dyn}$). Conventionally, a molecular gas fraction ($f_{\rm gas}$) is defined as $f_{\rm gas} = \frac{M_{\rm mol}}{M_{\rm dyn}}$. With this $f_{\rm gas}$, it indicates that $\frac{M_{\rm mol}}{M_{\rm dyn}}$ can be proportional to the ratio of $L'_{\rm CO}$ to FWHM$_{\rm CO}$, and this proportional relation can be formed as an identity as

$$f_{\rm gas} = \left(\frac{M_{\rm mol}}{M_{\rm dyn}}\right) = C \times \left(\frac{(L'_{\rm CO})^m}{(\text{FWHM}_{\rm CO})^n}\right). \quad (1)$$

Here the $C$, $m$, and $n$ are the parameters which can be obtained from observational data. Essentially, equation (1) can be regarded as a general expression of the Tully–Fisher relation (TFR, a relationship between the luminosity and rotational velocity of the nearby spiral galaxy; Tully & Fisher 1977). However, in the study of the CO LFR, while considering the morphological structures of the CO-detected galaxies (especially detected in $z > 1$) are not well-resolved in the current observations, the FWHM$_{\rm CO}$ has been used as a characteristic self-structural dynamical velocity of the CO-detected galaxy to replace the use of the rotational velocity. With the robust CO LFR, some publications also proposed to apply this scaling relation on the estimation of the magnification factor (denoted by $\mu$) of a gravitational-lensed galaxy (e.g. Harris et al. 2012; G15; Neri et al. 2020) or on the measurement of the cosmic distance of a galaxy in the high-$z$ Universe [e.g. G15; Wu et al. 2019 (hereafter W19); Wu 2019]. In G15, they verified that a smaller data scatter occurs in the CO LFR with the sole use of the CO(1-0) data.

★ E-mail: yhwastroph@xmu.edu.cn (Y-HW); jfwang@xmu.edu.cn (J-FW)
† Deceased 2022 May 21

Additionally, either in the current observations [e.g. Carilli & Walter 2013; Díaz-Santos et al. 2013, 2017; Farrah et al. 2013 (hereafter F13); Magdis et al. 2014 (hereafter M14); Ibar et al. 2015 (hereafter I15); Pozzi et al. 2021] or in the simulation (e.g. Vizgan et al. 2022), the emission line generated by the state transition of $^2P_{3/2}$ - $^2P_{1/2}$ of $C^+$ ([C II] 158 µm, or denoted as [C II] for convenience) has been also regarded as a reliable tracer to measure the galaxy characteristics. The strength of this [C II] emission is typically $>10^3$ times stronger than that of CO emission, implying the characteristic measured by the $C^+$ emission may lead to a smaller uncertainty than that measured by the CO emission. Motivated by the current observed data of the [C II]-detected galaxies and the previous studies about the CO LFR, especially the studies in G15 and W19 about the CO(1-0) LFR, we speculate that there might be also an similar relationship existing between the [C II] luminosity and FWHM linewidth (i.e. a [C II] LFR) in the [C II]-detected galaxy.

Therefore, in this paper, we investigate the possible existence of the [C II] LFR and examine the statistical properties (i.e. the data scatter and the redshift dependence) of the [C II] LFR. Moreover, we compare the analysis result of the [C II] LFR in this paper with that of the CO(1-0) LFR in W19. This paper is structured as the following: Section 2 shows the source of our sample and the selection, Section 3 describes the methodology,[1] Section 4 presents the result and discussion, Section 5 demonstrates the application of the [C II] LFR, and Section 6 concludes the findings in this paper and discusses possible future works about the [C II] LFR.

## 2 DATA

We selected the galaxies with [C II] measurements available in the literature. After the selection, we collected these selected galaxies into as a primary sample. To examine the redshift dependence of the [C II] LFR, we separated the primary sample into two subsamples according to whether the galaxy redshifts were smaller than one ($z < 1$, low $z$) or greater than one ($z > 1$, high $z$), and named the two subsamples as **low-$z$ [C II] sample** and **high-$z$ [C II] sample**. In Sections 2.1 and 2.2, we briefly state the data sources and selections.

### 2.1 Low-z sample

F13 published the observational results in *Herschel*. This reference paper provided the six fine-structure emission ([OIII]52 µm, [NIII]57 µm, [OI]63 µm, [NII]122 µm, [OI]145 µm, and [C II]158 µm) data in the 25 ultra-luminous infrared galaxies (ULIRGs) locating at $0.05 < z < 0.27$. We selected the 24 source objects from the 25 ULIRGs because their [C II] redshifts, [C II] fluxes (in $10^{-21}$ W cm$^{-2}$), and [C II] linewidths (in km s$^{-1}$, measured by fitting the [C II] emission lines with the single Gaussian profiles) were available in the reference paper. In Appendix, Table A1 lists the data of these 24 selected ULIRGs.

M14 published the results in the *Herschel*-SPIRE (Spectral and Photometric Imaging REceiver, Griffin et al. 2010) observation, and this reference paper presented the observed data of the 17 ULIRGs locating in $0.21 < z < 0.88$. We selected the seven galaxies of these 17 source galaxies because their redshift, [C II] fluxes, and [C II] FWHM linewidths were available in the paper. In Appendix, Table A2 lists the data of the seven selected galaxies.

I15 presented the results in the *Herschel*-PACS (Photodetector Array Camera and Spectrometer, Poglitsch et al. 2010) Integral Field Unit detection, and also published their observational results of the 28 [C II]-detected galaxies at $0.02 < z < 0.2$. We selected the 26 galaxies among the 28 [C II]-detected galaxies because of their available redshifts, [C II] fluxes, and [C II] FWHM linewidths in the reference paper. The data of the 26 selected galaxies are presented in Table A3 of Appendix.

### 2.2 High-z sample

Zanella et al. (2018, hereafter Z18) published the results of the *ALMA* Band 9 observation for the 10 main-sequence galaxies locating in $1.7 < z < 2.0$. We selected the four (Galaxy ID: 9347, 6515, 10076, and 9834) from the 10 main-sequence galaxies because of their available measured [C II] data (i.e. the [C II] redshifts, [C II] flux density $F_{\text{[CII]}}$, [C II] luminosity $L_{\text{[CII]}}$, and [C II] line velocity width $\Delta v$). These four galaxies as well as their measured [C II] data are presented in Table A4 of Appendix.

Decarli et al. (2018, hereafter D18) published their observed results of the 27 galaxies ($z > 6$, and they were identified as quasars) in the *ALMA* [C II] survey. The 23 galaxies of these 27 galaxies were selected because of their available [C II] redshifts, [C II] fluxes, and [C II] FWHM linewidths. In Appendix, Table A5 presents the data of these 23 selected galaxies.

In Cunningham et al. (2020, hereafter C20), they published the results of the 40 galaxies locating at $3 < z < 6$ in the *ACA* and *APEX* observations. The 30 galaxies of these 40 galaxies were selected because of their available [C II] redshifts, [C II] fluxes, and [C II] FWHM linewidths in this reference paper. In Appendix, Table A6 presents the data of these 30 selected galaxies.

A survey project named ALPINE (*ALMA* Large Program to Investigate $C^+$ at Early Time) aims to detect the [C II] emission at $4 < z < 6$. Moreover, a [C II] data set was released for the 118 galaxies observed in this survey project (Béthermin et al. 2020; Faisst et al. 2020; Le Fèvre et al. 2020). We checked the data set and then selected 75 galaxies because of their available [C II] redshifts, [C II] fluxes, and [C II] FWHM linewidths. In Appendix, Table A7 and A8 present the data of these selected 75 galaxies.

We also compiled several [C II]-detected galaxies from the following papers: Walter et al. (2009), Wagg et al. (2010a), Swinbank et al. (2012), Venemans et al. (2012, 2016), F13, Riechers et al. (2013), Wang et al. (2013, 2016, 2021), M14, I15, Bañados et al. (2015), Willott, Bergeron & Omont (2015a, 2017), Knudsen et al. (2016), Pavesi et al. (2016), Bradac et al. (2017), Neeleman et al. (2017), Umehata et al. (2017), D18, Smit et al. (2018), Z18, Wu (2019), C20, and Faisst et al. (2020). Among these papers, we selected 37 galaxies because their redshifts, [C II] fluxes, and [C II] FWHM linewidths were published and all available. The data of these 37 selected galaxies are presented in Table A9 of Appendix.

Through Sections 2.1 and 2.2, we eventually selected 226 [C II]-detected galaxies totally. The number distributions of our selected [C II]-detected sample galaxies in $z < 1$ and $z > 1$ are displayed in Fig. 1.

## 3 METHODOLOGY

Before investigating the correlation between the [C II] luminosity and FWHM linewidth, we processed these two quantities for our sample galaxies in both low- and high-$z$ [C II] samples. The methods of processing the data are stated in Sections 3.1 and 3.2.[2]

---

[1] Owing to comparing with the study in W19, we adopt the same spatially flat (curvature-free, $\Omega_k = 0$) $\Lambda$CDM model [$H_0 = 73.8$ km s$^{-1}$ Mpc$^{-1}$ in Riess et al. (2011), $\Omega_m = 0.295$ in Suzuki et al. (2012), and $(w_0, w_a) = (-1, 0)$] as that in W19.

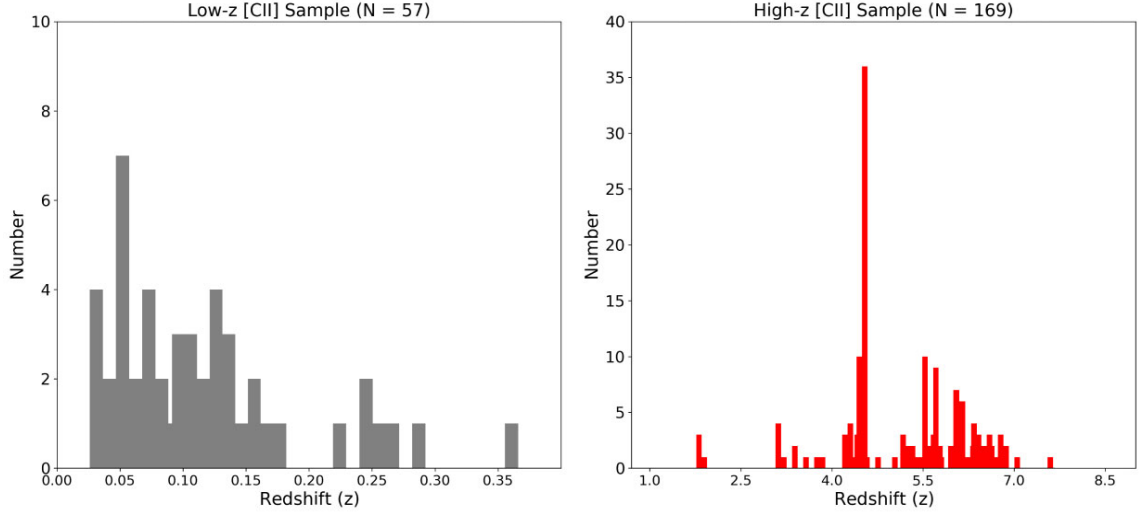

**Figure 1.** Left-hand panel: In low-$z$ ($z < 1$) [C II] sample, there are 57 sample galaxies spanning within $0.02 < z < 0.37$. Right-hand panel: In high-$z$ ($z > 1$) [C II] sample, there are 169 sample galaxies spanning within $1.85 < z < 7.65$.

### 3.1 [CII] luminosity

We derived the [C II] luminosity with the identity (Solomon et al. 1997; Emonts et al. 2011; Magnelli et al. 2012; Carilli & Walter 2013) which is expressed as

$$\frac{L'_{\rm [C\,II]}}{\rm [K\,km\,s^{-1}\,pc^2]} = \frac{3.25 \times 10^7 \, I_{[CII]} \, D_{\rm L}^2}{\nu_0^2 (1+z)}, \qquad (2)$$

where the $L'_{\rm [C\,II]}$ represents the [C II] luminosity in the unit of K km s$^{-1}$ pc$^2$, the $I_{\rm [CII]}$ represents the [C II] velocity-integrated flux (Jy km s$^{-1}$), and $\nu_0$ denotes the rest frequency of the [C II] emission ($\nu_0 = 1900.54$ GHz). In equation (2), the $D_{\rm L}$ and $z$ represent the luminosity distance (in Mpc) and redshift, respectively. We evaluated the $D_{\rm L}$ to the sample galaxy according to the mention in the footnote 1.

However, for the sample galaxies in F13 and M14, the units of their [C II] fluxes were not exactly in Jy km s$^{-1}$ but in either W m$^{-2}$ or W cm$^{-2}$. For this case, we adopted unit conversions that are expressed as

$$\begin{aligned}\frac{Q_1}{\rm [Jy\,km\,s^{-1}]} &= \left(\frac{3.0 \times 10^{22}}{\nu_0}\right) \times \left(\frac{Q_2}{\rm [W\,m^{-2}]}\right) \\ &= \left(\frac{3.0 \times 10^{26}}{\nu_0}\right) \times \left(\frac{Q_3}{\rm [W\,cm^{-2}]}\right),\end{aligned} \qquad (3)$$

where the units of the $Q_1$, $Q_2$, and $Q_3$ are in Jy km s$^{-1}$, W m$^{-2}$, and W cm$^{-2}$, respectively. Here $\nu_0$ is the rest frequency of the [C II] emission ($\nu_0 = 1900.537$ GHz). By equation (3), we converted the flux which is in either W m$^{-2}$ or W cm$^{-2}$ to the one in Jy km s$^{-1}$.

---

[2]Through checking the sample galaxies in high-$z$ [C II] sample (see Tables A4–A9), we noticed their redshift uncertainties are around three orders of magnitude smaller than the uncertainties of the measured [C II] flux and FWHM linewidth. Therefore, we considered that the redshift uncertainty has minimal impact on the uncertainty of the derivatives. Thus, in our data processing and analysis, for the sample galaxies with unavailable redshift uncertainties in the references, we assigned zeros to their redshift uncertainties (see Appendix).

An issue was concerned that the luminosity of a galaxy may be magnified due to the effect of gravitational lens. Therefore, the magnified luminosity should be corrected into an intrinsic one (Riechers 2011; Negrello et al. 2014) by

$$L'_{\rm [C\,II], intrins.} = \frac{L'_{\rm [C\,II]}}{\mu}, \qquad (4)$$

where the $\mu$ represents the magnification factor which is used to scale the degree of the magnification due to gravitational lens. Here, the $L'_{\rm [C\,II]}$ represents the derived [C II] luminosity and the $L'_{\rm [C\,II], intrins.}$ represents the intrinsic [C II] luminosity after the correction. We noticed that some of our high-$z$ ($z > 1$) sample galaxies had been observed in gravitational lens, indicating equation (4) should be applied to them. Here is a note that, in Section 2, for the high-$z$ sample galaxies observed in gravitational lens, their $\mu$ values had been provided in the publications and we also collected them into our data sets. Moreover, for the sample galaxies without being indicated their gravitational-lensed signs in the published papers, we assigned their $\mu$s as one (i.e. the sample galaxies in Tables A1, A2, A3, A4, A5, A7, and A8, and some sample galaxies in Table A9).

### 3.2 FWHM linewidth

For the four sample galaxies selected from Z18, the uncertainties of their line velocity widths were not given in this reference. To derive the uncertainties for these line velocity widths, we noticed the values listed in the column 8 ($F_{\rm [CII]}$, mJy) and column 9 ($L_{\rm [CII]}$, $10^9$ L$_\odot$) of the table 2 in Z18 had been available for deriving the uncertainty, and those values were also listed in the column 5 ($F_{\rm [CII]}$, mJy) and column 6 ($L_{\rm [CII]}$, $10^9$ L$_\odot$) of Table A4 (see Section 2.2). Thus, Table A4 can be referred when one is checking through **Step 1** and **Step 2** described next for the methods of the uncertainty derivation.

**Step 1:** We evaluated the velocity-integrated flux $I_{\rm [CII]}$ (Jy km s$^{-1}$) by using the value in the Column 6 of Table A4, and there is an identity (Carilli & Walter 2013) relating the [C II] flux ($I_{\rm [CII]}$, in

Jy km s$^{-1}$) to the [C II] luminosity ($L_{[CII]}$, in $L_\odot$) by

$$\frac{L_{[CII]}}{[L_\odot]} = \frac{1.04 \times 10^{-3} \times \nu_0 \times I_{[CII]} \times D_L^2}{(1+z)}, \quad (5)$$

where the $z$ denotes the redshift of the sample galaxy and the $\nu_0$ is the rest frequency of the [C II] emission ($\nu_0$ = 1900.537 GHz). The $D_L$ in equation (5) represents the luminosity distance in Mpc. However, the value of this $D_L$ here was determined based on the spatially flat Lambda cold dark matter (ΛCDM) cosmology model ($\Omega_m$ = 0.3, $\Omega_\Lambda$ = 0.7, and $H_0$ = 70.0 km s$^{-1}$ Mpc$^{-1}$) which had been adopted in Z18, and the value of this $D_L$ is also listed in the column 7 of Table A4. Conveniently, equation (5) can also be re-formed as

$$I_{[CII]} = \frac{L_{[CII]} \times (1+z)}{1.04 \times 10^{-3} \times \nu_0 \times D_L^2}. \quad (6)$$

With the given values of the $z$, $L_{[CII]}$, and $D_L$, the velocity-integrated flux $I_{[CII]}$ can be evaluated through equation (6). The measurement uncertainty $\sigma_{I_{[CII]}}$ of the evaluated $I_{[CII]}$ in equation (6) can be derived by

$$\sigma_{I_{[CII]}} = I_{[CII]} \times \sqrt{\left(\frac{\sigma_{L_{[CII]}}}{L_{[CII]}}\right)^2 + \left(\frac{\sigma_z}{(1+z)}\right)^2}, \quad (7)$$

where the $\sigma_{L_{[CII]}}$ and $\sigma_z$, respectively, represent the measurement uncertainties of the $L_{[CII]}$ (in $L_\odot$) and redshift $z$. Herein, the $I_{[CII]}$ as well as the $\sigma_{I_{[CII]}}$ evaluated from equation (6) and (7) can also be taken as the velocity-integrated [C II] flux to the selected Zanella sample galaxy if equation (2) is used.

**Step 2:** The velocity-integrated flux $I_{[CII]}$ itself is also a quantity that can be derived by integrating with a single Gaussian model (which is applied to fit a [C II]-line profile) in a velocity interval. Thus, mathematically, there is a relationship existing among the velocity-integrated $I_{[CII]}$, the [C II] flux density $F_{[CII]}$ (if one regards it as an amplitude of the applied single Gaussian model), and the FWHM linewidth of the applied single Gaussian profile, and this relationship can be expressed as

$$\frac{I_{[CII]}}{[\text{Jy km s}^{-1}]} = \left(\frac{1000\sqrt{2\pi}}{2\sqrt{2\ln(2)}}\right) \times \left(\frac{F_{[CII]}}{[\text{mJy}]}\right) \times \left(\frac{\text{FWHM}}{[\text{km s}^{-1}]}\right), \quad (8)$$

or it can be re-formed as

$$\text{FWHM} = \left(\frac{2\sqrt{2\ln(2)}}{1000\sqrt{2\pi}}\right) \times \left(\frac{I_{[CII]}}{F_{[CII]}}\right). \quad (9)$$

By checking the methodologies used in Z18, we found that they also utilized single Gaussian profiles in their [C II]-line fits. Therefore, by equation (9), one is able to derive the FWHM linewidth with the evaluated $I_{[CII]}$ in equation (6) and the value of $F_{[CII]}$ listed in the column 5 of Table A4. Besides, the measurement uncertainty $\sigma_{\text{FWHM}}$ of the FWHM in equation (9) can be computed by

$$\sigma_{\text{FWHM}} = \text{FWHM} \times \sqrt{\left(\frac{\sigma_{I_{[CII]}}}{I_{[CII]}}\right)^2 + \left(\frac{\sigma_{F_{[CII]}}}{F_{[CII]}}\right)^2}, \quad (10)$$

where the FWHM was derived from equation (9), and the $\sigma_{I_{[CII]}}$ and $\sigma_{F_{[CII]}}$ are the measurement uncertainties of the $I_{[CII]}$ and $F_{[CII]}$, respectively.

By the above two steps, we derived the FWHM linewidths as well as the uncertainties to the four sample galaxies selected from Z18.

We also concerned the influence of the dynamical structure of a galaxy on the FWHM linewidth measured in the [C II]-line profile. For an isolated spiral galaxy with its inclination ($i$, in degree), its FWHM linewidth measured in the [C II] line is a projected quantity due to the galaxy inclination. To obtain an intrinsic FWHM linewidth, the measured FWHM linewidth must be corrected through dividing by a factor of sin ($i$) (e.g. Tiley et al. 2016; Topal et al. 2018; Kohandel et al. 2019) which is defined as

$$\text{FWHM}_{[CII]}^{\text{intris.}} = \frac{\text{FWHM}_{[CII]}}{\sin(i)}. \quad (11)$$

Here, the $i$ and sin ($i$) represent the galaxy inclination ($i$) and correction factor, respectively. The FWHM$_{[CII]}^{\text{intris.}}$ represents the intrinsic [C II] FWHM linewidth after the correction and the FWHM$_{[CII]}$ represents the measured [C II] FWHM linewidth before the correction. After reviewing the sample galaxies in low-$z$ [C II] sample, some of them were published with their well-resolved spiral-disc features (including their observed inclinations), indicating their observed FWHM$_{[CII]}$ should be corrected by equation (11). However, for the rest of the sample galaxies in low-$z$ [C II] sample or even the sample galaxies in high-$z$ [C II] sample, their morphological structures had not been well-resolved, not to mention any structural corrections for their observed FWHM linewidth, although their observed FWHM linewidths were analysed. After referring to the methodology in W19, for both low- and high-$z$ [C II] samples, we decided to directly take their observed FWHM linewidths in the investigation of the significance of the [C II] LFR.

### 3.3 Linear regression

We set up a method for investigating the significance of the [C II] LFR in low-$z$ [C II] sample as well as the one in high-$z$ [C II] sample. For an individual [C II] sample (i.e. either low- or high-$z$ [C II] sample), the data of the $\log_{10}(L'_{[CII]})$ and $\log_{10}(\text{FWHM}_{[CII]})$ were depicted in a $\log_{10}(L'_{[CII]})$–$\log_{10}(\text{FWHM}_{[CII]})$ plane. To investigate the linear trend possibly existing among the data depicted in the $\log_{10}(L'_{[CII]})$–$\log_{10}(\text{FWHM}_{[CII]})$ plane, we assumed a linear model (a straight line) which can best fit to the data, and the properties (i.e. the slope and intercept, and IS that represent the data scatter in the best-fitting model line) of this assumed linear model were derived by implementing the LR algorithm. Owing to comparing with the result in W19, on setting up our assumed linear model, we referred to the methodology in W19 to consider a pivot point and assign the $\log_{10}(L'_{[CII]})$ and $\log_{10}(\text{FWHM}_{[CII]})$ of the [C II] sample as the independent and dependent variables, respectively. Thus, the assumed linear model line can be formed as

$$\log_{10}(\text{FWHM}_{[CII]}) - A = \alpha + \beta \left(\log_{10}(L'_{[CII]}) - B\right). \quad (12)$$

Here, the $\alpha$ and $\beta$ represent the intercept and slope which can be derived by performing with the Python package called **linmix**,[3] and the $A$ and $B$ are in pair to denote the pivot point of equation (12) in the $\log_{10}(L'_{[CII]})$–$\log_{10}(\text{FWHM}_{[CII]})$ plane.

The pair of the $A$ and the $B$ are used to indicate the location where the [C II] sample data gather concentratedly more in the $\log_{10}(L'_{[CII]})$–$\log_{10}(\text{FWHM}_{[CII]})$ plane. To assign the pivot point for the linear model line, by again referring to the methods in W19, we combined low- and high-$z$ [C II] samples as a single data set named **combined sample** and then computed the median values of the $\log_{10}(L'_{[CII]})$ and $\log_{10}(\text{FWHM}_{[CII]})$ in combined sample. By this computation, we obtained $A$ = 2.489 and $B$ = 9.480, and assigned them to be as the pivot point of low-$z$ [C II]. Similarly, we also assigned these same values to be as the pivot point of high-$z$ [C II]. Thus, being as one case, low-$z$ [C II] sample and high-$z$ [C II] sample use the same pivot point (i.e. the pivot point of combined sample) in their own LR algorithms.

---

[3]Kelly (2007) constructed this Python package whose LR algorithm takes the measurement errors of the independent and dependent variables into account. Moreover, the LR algorithm in the package also derives the IS about the best-fitting linear model.

**Table 1.** The results from the linear regression (LR) analyses. For the case of using the pivot point of combined sample (low $z$ [C II] + high $z$ [C II]), the LR analysis results for low- and high-$z$ [C II] samples are, respectively, presented in the second and third rows. Similarly, for the case of using their own pivot points, the LR analysis results for low- and high-$z$ [C II] samples are, respectively, presented in the fourth and fifth rows. The LR analysis result for combined sample is presented in the sixth row. The first column presents the name of the sample. The second column presents the number ($N$) of the galaxy in the sample. The third column presents the pivot point ($A$, $B$). The fourth and fifth columns present the best-fitting values of the intercept ($\alpha$) and slope ($\beta$), respectively. The sixth column presents the value of the intrinsic scatter (IS, $\epsilon_i$) which scales the sample data scattering around the best-fitting linear model line. In the fourth, fifth, and sixth columns, the values located behind the $\pm$ signs indicate the $1\sigma$ uncertainties. The seventh column presents the Spearman's coefficient (SC) which indicates the degree of being a power-law correlation between the $\log_{10}(L'_{\text{[C II]}})$ and $\log_{10}(\text{FWHM})$. The eighth column presents the corresponding $p$-value to the SC presented in the seventh column, and a criterion that $p < 0.05$ indicates an appearance of a significant $\log_{10}(L'_{\text{[C II]}})$ and $\log_{10}(\text{FWHM})$ correlation. The ninth column presents the value of $\rho$ which manifests the covariance Cov.($\alpha$, $\beta$) of the resulting intercept ($\alpha$) and slope ($\beta$) values in the LR algorithm.

| [C II] sample | $N$ | ($A$, $B$) | $\alpha$ | $\beta$ | IS ($\epsilon_i$) | SC | $p$-value | $\rho = $ Cov.($\alpha$, $\beta$) |
|---|---|---|---|---|---|---|---|---|
| Low $z$ | 57 | (2.489, 9.480) | $0.049 \pm 0.028$ | $0.218 \pm 0.056$ | $0.168 \pm 0.018$ | 0.443 | 0.0006 | 0.578 |
| High $z$ | 169 | (2.489, 9.480) | $-0.028 \pm 0.015$ | $0.213 \pm 0.027$ | $0.170 \pm 0.012$ | 0.457 | $4.394 \times 10^{-10}$ | $-0.235$ |
| Low $z$ | 57 | (2.471, 9.267) | $0.020 \pm 0.023$ | $0.218 \pm 0.055$ | $0.167 \pm 0.018$ | 0.443 | 0.0006 | 0.182 |
| High $z$ | 169 | (2.504, 9.595) | $-0.018 \pm 0.014$ | $0.213 \pm 0.028$ | $0.170 \pm 0.012$ | 0.457 | $4.394 \times 10^{-10}$ | $-0.015$ |
| Combined | 226 | (2.489, 9.480) | $-0.007 \pm 0.012$ | $0.191 \pm 0.023$ | $0.170 \pm 0.009$ | 0.448 | $1.538 \times 10^{-12}$ | $-0.017$ |

By the method of computing the pivot point mentioned previously, we also computed the pivot points to low-$z$ [C II] sample and high-$z$ [C II] sample, separately. That is, for low-$z$ [C II] sample, the median values of 2.471 and 9.267 were computed and assigned to the $A$ and $B$, respectively, similarly, for high-$z$ sample, the median values of 2.504 and 9.595 were assigned to the $A$ and $B$. Thus, being another case in our LR algorithm, low- and high-$z$ [C II] sample use their own pivot points in their LR algorithms.

In short, two cases, 'using the pivot point of combined sample' and 'using their own pivot points', were labelled and arranged for our work of the LR analysis. For each case, there were two [C II] samples (low- and high-$z$ [C II] sample), and each [C II] sample was taken to the LR algorithm.

## 4 RESULTS AND DISCUSSIONS

Table 1 lists the results in our LR analyses, and Fig. 2 depicts the results listed in Table 1 with the data of low- and high-$z$ [C II] samples in the $\log_{10}(L'_{\text{[C II]}})$–$\log_{10}(\text{FWHM}_{\text{[C II]}})$ planes. The caption of Fig. 2 states the meanings of denotations adopted in the depictions. Moreover, the LR results of combined sample are presented in Table 1 and Fig. 2 for comparison.

For the case of using the pivot point of combined sample (in the second and third rows of Table 1 and the left-hand panel of Fig. 2). We found that, in both low- and high-$z$ [C II] sample, their SCs are in $0.4 < \text{SC} < 0.5$ and their $p$-values are smaller than the threshold of 0.05. It indicates that the [C II] LFR exists robustly in low-$z$ [C II] sample and it also does in high-$z$ [C II] sample. The slope of the low-$z$ [C II] LFR line is closely consistent with that of the high-$z$ [C II] LFR line in the $1\sigma$ level if one checks the contour plot in the top-left corner of the left-hand panel of Fig. 2. However, the intercepts of the low- and high-$z$ [C II] LFR lines locate out of the $3\sigma$ levels away from each other, manifesting an obvious intercept discrepancy in the two [C II] LFR lines. By comparing the IS results in this case with those in W19 (i.e. the IS result in the low-$z$ CO(1-0) sample is $0.20 \pm 0.02$ and the one in the high-$z$ CO(1-0) sample is $0.20 \pm 0.03$, and the numbers placed behind the $\pm$ signs are the $1\sigma$ uncertainties), the IS result in low-$z$ [C II] sample reduces by $0.032 \pm 0.027$ with respect to the one in the low-$z$ CO(1-0) sample and the IS result in high-$z$ [C II] sample reduces by $0.030 \pm 0.032$ with respect to the one in the high-$z$ CO(1-0) sample. If we involved their uncertainties in the comparison, for the low-$z$ CO(1-0) sample versus low-$z$ [C II] sample, the reduction of 0.032 is slightly over the uncertainty range of 0.03 ($\sqrt{0.02^2 + 0.018^2}$), and, for the high-$z$ CO(1-0) sample versus high-$z$ [C II] sample, the reduction of 0.030 is exactly at the uncertainty range of 0.03 ($\sqrt{0.03^2 + 0.012^2}$). Thus, for the samples in either $z < 1$ (low-$z$) or $z > 1$ (high-$z$), we may conclude the strength of the [C II] data scatter about the [C II] LFR corresponds to that of the CO(1-0) data scatter about the CO(1-0) LFR.

For the case of using their own pivot points (in the fourth and fifth rows of Table 1 and the right-hand panel of Fig. 2), as similar findings as those mentioned in the previous case, there is a [C II] LFR existing in a robust level ($0.4 < \text{SC} < 0.5$ and $p < 0.05$) in low-$z$ [C II] sample and it does in high-$z$ [C II] sample. And, the slope of the low-$z$ [C II] LFR line is closely consistent with that of the high-$z$ one in the $1\sigma$ level. The intercept discrepancy between the two [C II] LFRs still occurs but this discrepancy becomes reduced in this case. Moreover, both the IS values in low- and high-$z$ [C II] samples in this case changes a few with respect to those in low- and high-$z$ [C II] samples in the previous case. Therefore, while we compared the IS results in this case with the IS results in W19, we obtained the same conclusions which had been mentioned in the previous case.

For the $\rho$ values (which indicate the correlation between the resulting slope and intercept in the LR algorithm) presented in the ninth column of Table 1, no matter which the case is, a positive correlation ($\rho > 0$) is manifest in low-$z$ [C II] sample and a negative correlation ($\rho < 0$) in high-$z$ [C II] sample. Moreover, as taking absolute value in the $\rho$ values ($|\rho|$s), we found both the $|\rho|$s of low- and high-$z$ [C II] sample in the case of using their own pivot points become dropped with respect to those $|\rho|$s of low- and high-$z$ [C II] sample in the case of using the pivot point of combined sample.

From the previous two cases, whatever the pivot point of combined sample is used or not, there are similar findings about the [C II] LFRs in low- and high-$z$ [C II] sample. We summarized the findings as the following points: (1) The [C II] LFR exists significantly; (2) the slope of the [C II] LFR presents no strong evolution in redshift; however, the intercept of the [C II] LFR presents a shift while redshift increases; (3) the absolute value of the $\rho$ becomes reduced for the case of using their own pivot points and the IS value remains unchanged for both two cases; and (4) while comparing with the IS result in W19, the degree of the data scatter about the [C II] LFR corresponds to that about the CO(1-0) LFR.

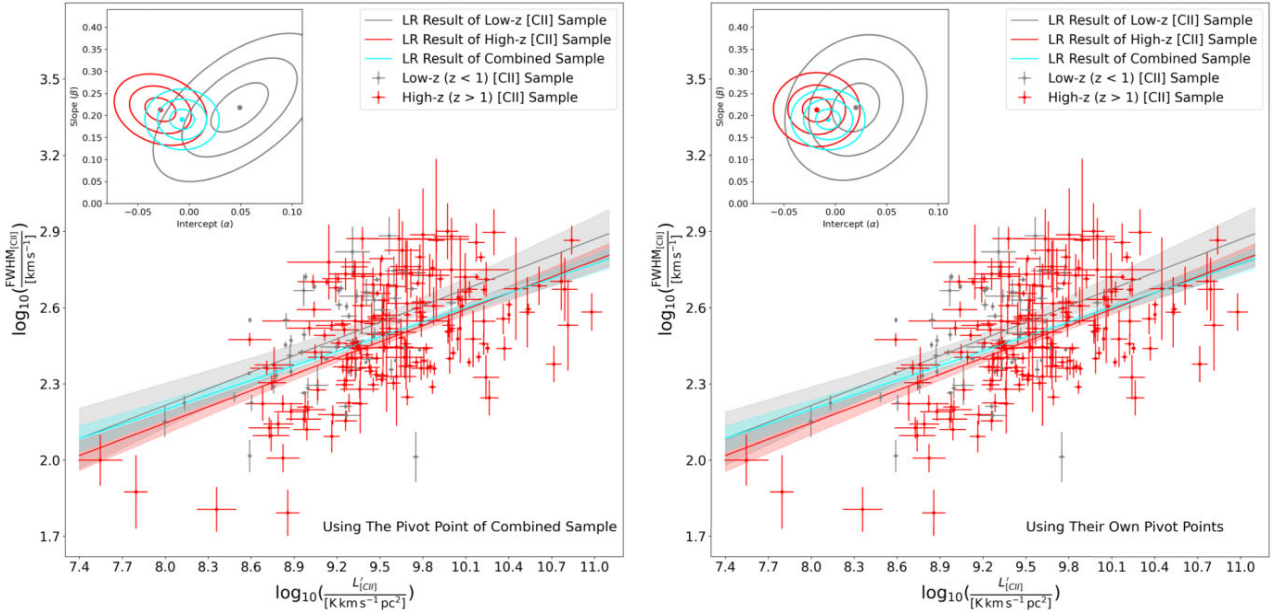

**Figure 2.** Left-hand panel: Using the pivot point ($A = 2.489$ and $B = 9.480$) of combined sample. In this panel, the grey straight line represents the best-fitting LR model line to low-$z$ [C II] sample (the grey dots) with the light-grey area which represents the $1\sigma$ uncertainty field around the grey best-fitting model line. Similarly, the red straight line represents the best-fitting LR model line about high-$z$ [C II] sample (the red dots) with the light-red area which represents the $1\sigma$ uncertainty field around the red best-fitting model line. The cyan line represents the best-fitting LR model line about combined sample (low- + high-$z$ [C II] samples) with the cyan area which represents the $1\sigma$ uncertainty field around the cyan best-fitting model line. For each sample data dot, the vertical and horizontal bars represent the $1\sigma$ measurement uncertainties of the $\log_{10}$(FWHM) and $\log_{10}(L'_{[CII]})$, respectively. In the left-top corner of this panel, this contour plot displays the outcome distributions of the slope ($\beta$) and intercept ($\alpha$) in the LR algorithms with the [C II] samples. The grey contours illustrate the confidence levels (the inner, middle, and outer contours correspond to the $1\sigma$, $2\sigma$, and $3\sigma$ levels) of the outcome distribution of the slope ($\beta$) and intercept ($\alpha$) in the two-dimensional plane (the $\alpha$–$\beta$ plane) for low-$z$ [C II] sample, and the red contours and cyan contours illustrate these for the high-$z$ [C II] sample and combined sample, respectively. In the contour plots, the grey dot located in the centre of the grey contours represents the best-fitting values of the slope ($\beta$) and intercept ($\alpha$) in the LR algorithm with low-$z$ [C II] sample, the red dot represents the best-fitting values of the slope ($\beta$) and intercept ($\alpha$) with high-$z$ [C II] sample, and the cyan dot represents the best-fitting values of the slope ($\beta$) and intercept ($\alpha$) with combined sample. Right-hand panel: Using their own pivot points. ($A$, $B$) = (2.471, 9.267) was used in the LR algorithm with low-$z$ [C II] sample, ($A$, $B$) = (2.504, 9.595) with high-$z$ [C II] sample, and ($A$, $B$) = (2.489, 9.480) with combined sample. In this right-hand panel, all the denotations are as the same as those adopted in the left-hand panel.

## 5 POSSIBLE APPLICATIONS

### 5.1 Measurement of cosmic distance

In Section 3.1, it was feasible to derive the luminosity distance $D_L$ to the sample galaxy by using the $\Lambda$CDM model as described in Section 3. However, an issue is here that if a distance indicator (an object whose distance is well-measured and reliable in observation) replaces the usage of the $\Lambda$CDM model (or other cosmological models) to calibrate the cosmic distance of a galaxy, whether the [C II] LFR itself, like as the astronomical distance ladders [e.g. the eclipsing binaries, the Cepheid variables, and the Type-Ia supernovae (SNe Ia, Kaluzny et al. 1998; Willick 1999)], could be a tool to measure the cosmic distance in $z > 1$ through using the statistical properties found in Section 4. In this subsection, we demonstrate for this issue step by step under no consideration of any cosmological model.

#### 5.1.1 Step 1. The distance calibration

The Union 2.1 compilation (Suzuki et al. 2012) of the Supernova Cosmology Project collected 580 SNe Ia. In this compilation, 580 SNe Ia are listed with their source IDs, redshifts, reliable-measured distance moduli (DMs) as well as the $1\sigma$ DM errors, and the probabilities of the SNe existing in a low-mass galaxy. Fig. 3 depicts the DM values in the Union 2.1 sample as a function of redshift. Apparently, the SNe Ia sample data denoted by the red dots manifest a linear correlation between in the DM–$\log_{10}$(z) plane. By assuming this linear correlation with a liner model, we derived the best-fitting slope and intercept of the assumed linear model with the Python code **linmix**, and the best-fitting results of the assumed linear model are shown as

$$\text{DM} = (43.913 \pm 0.016) + (5.495 \pm 0.016) \times \log_{10}(z). \quad (13)$$

The best-fitting slope and intercept with their $1\sigma$ uncertainties are $5.495 \pm 0.016$ and $43.913 \pm 0.016$, respectively, with the IS of 0.089, the SC of 0.993, and the $\rho$ value of 0.835. With all information about equation (13), the corresponding uncertainty $\sigma_{DM}$ of the DM in equation (13) was derived by

$$\sigma_{DM}^2 = 0.016^2 + (\log_{10}(z) \times 0.016)^2 + 0.089^2 + \left(\frac{5.495\,\sigma_z}{z\ln(10)}\right)^2$$
$$+ 2 \times 0.835 \times 0.016 \times 0.016 \times \log_{10}(z). \quad (14)$$

Through equations (13) and (14), and the redshift value in low-$z$ [C II] sample, we measured the DM as well as its uncertainty $\sigma_{DM}$ to the galaxy in low-$z$ [C II] sample. Afterwards, we converted the measured DM into a distance $D$ in Mpc through

$$\frac{D}{[\text{Mpc}]} = 10^{\frac{\text{DM}-25}{5}}. \quad (15)$$

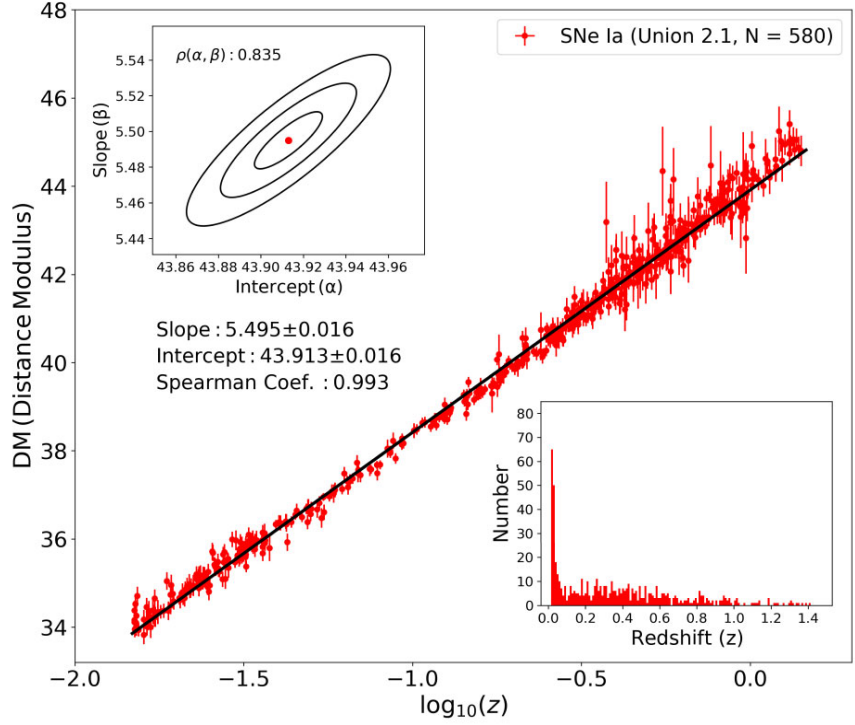

**Figure 3.** The red dots represent the DMs of the SNe Ia in the Union 2.1 (Suzuki et al. 2012) and the vertical bars represent the measurement uncertainties of the DMs. The black line represents the best-fitting linear trend of the SNe Ia data, with the best-fitting slope of $5.495 \pm 0.016$ and the best-fitting intercept of $43.913 \pm 0.016$. The intrinsic data scatter in the best-fitting black line is 0.008 and the SC is 0.993. Upper-left panel: This upper-left panel illustrates the confidence-level contour for the Union 2.1 sample in the slope ($\beta$)–intercept ($\alpha$) plane. The central red dot represents the best-fitting values of the intercept and slope, and the inner, middle, and outer rings represent the $1\sigma$, $2\sigma$, and $3\sigma$ confidence levels, respectively. The covariance $\rho(\alpha, \beta)$ of the possible $\alpha$ and $\beta$ values is 0.835. This resulting contour was derived by the Python package **linmix**. Lower-right panel: This lower-right panel displays the number distribution of the Union 2.1 SNe Ia sample in the whole-compilation redshift range of $0.015 < z < 1.414$.

Here, in equation (15), we calibrated the $D$ to the galaxy in low-$z$ [C II] sample. Accordingly, the uncertainty $\sigma_D$ of the calibrated $D$ from equation (15) were derived by

$$\sigma_{D_L} = \left(\frac{\ln(10)}{5}\right) D_L \sigma_{DM}. \quad (16)$$

*5.1.2 Step 2. The calibration of the [C II] LFR in $z < 1$*

With the outcome of the $D$ in equation (15) and the definition of equation (2), we were able to calibrate the [C II] luminosity to the galaxy in low-$z$ [C II] sample. Later, by executing the **linmix** with the calibrated [C II] luminosity and given FWHM$_{[C II]}$ of the galaxy in low-$z$ [C II] sample, we obtained a calibrated low-$z$ [C II] LFR that is presented as

$$\log_{10}(\text{FWHM}_{[C II]}) - 2.471 = (0.020 \pm 0.023)$$
$$+ (0.210 \pm 0.054)$$
$$\times (\log_{10}(L'_{[C II]}) - 9.347), \quad (17)$$

where the $0.210 \pm 0.054$ and $0.020 \pm 0.023$ are the best-fitting slope ($\beta$) and intercept ($\alpha$), respectively, with their own $1\sigma$ uncertainties. Here, the number pair of (2.471, 9.347) is the pivot point. The IS $\epsilon_i$ and the SC are 0.167 and 0.435, respectively, with the corresponding $p$-value of 0.0007. The covariance $\rho$ between the $\alpha$ and $\beta$ is 0.194. The slope and intercept of equation (17) are consistent with those listed in the fourth row of Table 1, confirming the [C II] LFR itself is a robust existence.

*5.1.3 Step 3. The application of the calibrated low-$z$ [C II] LFR*

In Step 2, the pivot point of (2.471, 9.347) had been determined with the sole use of low-$z$ [C II] sample, and this number pair was used as the pivot point on obtaining equation (17). This situation is actually equivalent to the case of using their own pivot points mentioned in Section 3.3 and the relevant results of this case were presented in the right-hand panel of Fig. 2. Checking the right-hand panel of Fig. 2, the slope of the high-$z$ [C II] LFR is deviated from that of the low-$z$ [C II] LFR in the $1\sigma$ level, and the intercept of the high-$z$ [C II] LFR is deviated from that of the low-$z$ [C II] LFR in the $2\sigma$ level. It implies the slope and intercept of the high-$z$ [C II] LFR can be regarded as similarly close as these of the low-$z$ [C II] LFR. That is, the [C II] LFR in this situation can be regarded as a redshift-independent scaling relation, and it may inspire that the [C II] LFR determined in $z < 1$ can be applied to the galaxy in high-$z$ [C II] sample. With this redshift independence of the [C II] LFR we derived, equation (17) can be utilized to measure the [C II] luminosity to the galaxy in high-$z$ [C II] sample. Conveniently, equation (17) can be re-formed as

$$\log_{10}(L'_{[C II]}) = \left(\frac{\log_{10}(\text{FWHM}_{[C II]}) - 2.471 - 0.020}{0.210}\right) + 9.347, \quad (18)$$

where the $\log_{10}(L'_{[C II]})$ was obtained with the given observed FWHM linewidth of the galaxy in high-$z$ [C II] sample. Correspondingly, the uncertainty $\sigma_{\log_{10}(L'_{[C II]})}$ of the $\log_{10}(L'_{[C II]})$ in equation (18) can be

derived by

$$\sigma^2_{\log_{10}(L'_{[\text{C II}]})} = \left(\frac{\sigma_{\log_{10}(\text{FWHM}_{[\text{C II}]})}}{0.210}\right)^2 + \left(\frac{0.023}{0.210}\right)^2$$
$$+ \left(\frac{(\log_{10}(\text{FWHM}_{[\text{C II}]}) - 2.471 - 0.020) \times 0.054}{0.210^2}\right)^2$$
$$+ \left(\frac{2 \times (\log_{10}(\text{FWHM}_{[\text{C II}]}) - 2.471 - 0.020)\sigma_{\alpha\beta}}{0.210^3}\right)$$
$$+ 0.028, \quad (19)$$

where the IS (0.167) of equation (17) is also considered in this uncertainty calculation ($0.028 = 0.167^2$). The term $\sigma_{\alpha\beta}$ is defined as

$$\sigma_{\alpha\beta} = \rho \sigma_\alpha \sigma_\beta, \quad (20)$$

where the $\rho$, $\sigma_\alpha$, and $\sigma_\beta$ are 0.194, 0.023, and 0.054, respectively.

By equation (18) and 19, we derived the $\log_{10}(L'_{[\text{C II}]})$ and the corresponding uncertainty $\sigma_{\log_{10}(L'_{[\text{C II}]})}$ to the galaxy in high-$z$ [C II] sample. The derived $\log_{10}(L'_{[\text{C II}]})$ can be converted to $L'_{[\text{C II}]}$ through a simply mathematical relation

$$L'_{[\text{C II}]} = 10^{\log_{10}(L'_{[\text{C II}]})}, \quad (21)$$

and the uncertainty $\sigma_{L'_{[\text{C II}]}}$ of the $L'_{[\text{C II}]}$ can be as the below:

$$\sigma_{L'_{[\text{C II}]}} = L'_{[\text{C II}]} \times \ln(10) \times \sigma_{\log_{10}(L'_{[\text{C II}]})}. \quad (22)$$

Through equation (21) and (22), we derived the [C II] luminosity as well as its uncertainty to the galaxy in high-$z$ [C II] sample.

*5.1.4 Step 4. The derivation of the cosmic distance in $z > 1$*

After the outcomes in equation (21) and (22), in this step, we start to compute the luminosity distance $D_\text{L}$ to the galaxy in high-$z$ [C II] sample by using equation (2). For convenience, equation (2) can be re-formed as

$$\frac{D_\text{L}}{[\text{Mpc}]} = \sqrt{\frac{v_0^2 \times (1 + z) \times \mu \times L'_{[\text{C II}]}}{3.25 \times 10^7 I_{[\text{CII}]}}}. \quad (23)$$

Here, the $I_{[\text{CII}]}$, $\mu$, and $z$ were already given in high-$z$ sample and the $L'_{[\text{C II}]}$ was the derived quantity from equation (21). Thus, we were able to compute the $D_\text{L}$ to the galaxy in high-$z$ [C II] sample by equation (23). The corresponding uncertainty $\sigma_{D_\text{L}}$ of the $D_\text{L}$ in equation (23) was computed by

$$\sigma^2_{D_\text{L}} = \left(\frac{D_\text{L}}{2}\right)^2$$
$$\times \left[\left(\frac{\sigma_{I_{[\text{CII}]}}}{I_{[\text{CII}]}}\right)^2 + \left(\frac{\sigma_{L'_{[\text{C II}]}}}{L'_{[\text{C II}]}}\right)^2 + \left(\frac{\sigma_\mu}{\mu}\right)^2 + \left(\frac{\sigma_z}{1+z}\right)^2\right]. \quad (24)$$

Here, the $\sigma_{I_{[\text{CII}]}}$, $\sigma_{L'_{[\text{C II}]}}$, $\sigma_\mu$, and $\sigma_z$ denote the uncertainties of the $I_{[\text{CII}]}$, $L'_{[\text{C II}]}$, $\mu$, and $z$, respectively. In convention, the computed $D_\text{L}$ in equation (23) can also be converted to DM by

$$\text{DM} = 25 + 5 \times \log_{10}\left(\frac{D_\text{L}}{[\text{Mpc}]}\right). \quad (25)$$

The uncertainty $\sigma_\text{DM}$ of the DM in equation (25) was derived from the uncertainty $\sigma_{D_\text{L}}$ in equation (24), which can be related by an expression as

$$\sigma_\text{DM} = \left(\frac{5}{\ln(10)}\right) \times \left(\frac{\sigma_{D_\text{L}}}{D_\text{L}}\right). \quad (26)$$

Through the previous four steps, we observationally measured the cosmic distances (or DMs) to the galaxies in high-$z$ [C II] sample, and their resulting measured DMs were also depicted in Fig. 4. In Fig. 4, the red, grey, and cyan dots denote the measured observed DMs of high-$z$ [C II] sample, the Union 2.1 SNe sample, and the high-$z$ CO ($J$ = 1-0) sample, respectively. The DM of the high-$z$ CO ($J$ = 1-0) sample referred to W19. For comparison, in Fig. 4, we also displayed the spatially flat concordance ΛCDM model depicted as the black curve.

In Fig. 4, with the DMs of the [C II] sample galaxies in $z > 1$, the cosmic spatial scale is extended to the early epoch ($z > 4$) of the Universe, and, by comparing with the ΛCDM model ($w_0 = -1$ and $w_a = 0$) demonstrated by the black curve, it is apparent that these three measured DM samples, i.e. the Union 2.1, high-$z$ [C II] sample, and the high-$z$ CO(1-0) sample, seem manifesting an evolutionary cosmic spatial growing trend along different cosmic redshifts. This cosmic growing trend is exactly manifesting the expansion of the Universe and those samples are believed to follow the rate of this cosmic expansion. In other words, the DM values of the three samples can be regarded as a tracer of marking the rate of the cosmic expansion. So far, it is believed that the time-varying dark energy drives the cosmic expansion in a rate of acceleration by the observations of the SNe Ia (Riess et al. 1998; Schmidt et al. 1998; Perlmutter et al. 1999). Because of the time-varying nature of dark energy, both the $w_0$ and $w_a$, which are sensitive to the nature of dark energy (King et al. 2014; Lukovic, Haridasu & Vittorio 2018), are indicated no longer to be −1 and 0, respectively, since these two values imply no existence of the time-varying dark energy. Therefore, it is expected that the evolutionary cosmic trend manifested by the three samples should be distinct from that expected by the ΛCDM model curve, and we may regard the DMs in Fig. 4 as a tracer to manifest the time-varying nature of dark energy. With this idea, we modelled for dark energy with the two cosmological parameters, the $w_0$ and the $w_a$, and constrained these two parameters with the measured DMs in Fig. 4. The next subsection presents the work for this cosmological parameter constraint.

### 5.2 Cosmological parameter constraint

Before the cosmological parameter constraint, we constructed a theoretical framework that considers the evolution of dark energy. Theoretically, the luminosity distance $D_\text{L}$ of a galaxy locating at a redshift $z$ can be expressed as (Hogg 1999; Signore & Puy 2001):

$$\frac{D_\text{L}}{[\text{Mpc}]} = \frac{c(1+z)}{H_0} \int_0^z \frac{dz'}{E(z', \mathbf{p})}. \quad (27)$$

Here, the $c$ and $H_0$ are the speed of light ($3 \times 10^5$ km s$^{-1}$) and present-day Hubble constant in km s$^{-1}$ Mpc$^{-1}$, respectively. The prime ' notated in the superscript of the redshift $z$ indicates that the $E(z', \mathbf{p})$ function is integrated in the redshift interval from the zero to the specific redshift value $z$. And, the $E(z, \mathbf{p})$ itself is a function of redshift $z$ and of a cosmological-parameter set $\mathbf{p}$ [including the matter density $\Omega_\text{m}$, the dark energy density $\Omega_\text{DE}$, the curvature density $\Omega_\text{k}$, and the equation of state (EoS) of dark energy $w$].

In physics, the EoS of dark energy $w$ is defined as a ratio of dark-energy pressure $P_\text{DE}$ to the dark-energy density $\rho_\text{DE}$ (Montiel & Bretón 2011; Lukovic et al. 2018), which can be expressed as

$$w = \frac{P_\text{DE}}{\rho_\text{DE}}. \quad (28)$$

In case of the time-varying nature of dark energy (i.e. dark energy evolving along different cosmic redshifts), the $w$ itself can be regarded as a function of redshift $w(z)$, and correspondingly, the $E(z, \mathbf{p})$ can be transferred into the form by, e.g. Montiel & Bretón 2011, King et al. 2014, and Lukovic et al. 2018, as

$$E(z, \Omega_m, \Omega_{DE}, \Omega_k) = \sqrt{\Omega_m(1+z)^3 + \Omega_k(1+z)^2 + \Omega_{DE}f(z)}, \quad (29)$$

where the $f(z)$ is defined as a form of

$$f(z) = \exp\left(3 \int_0^z \frac{1+w(z')}{1+z'} dz'\right). \quad (30)$$

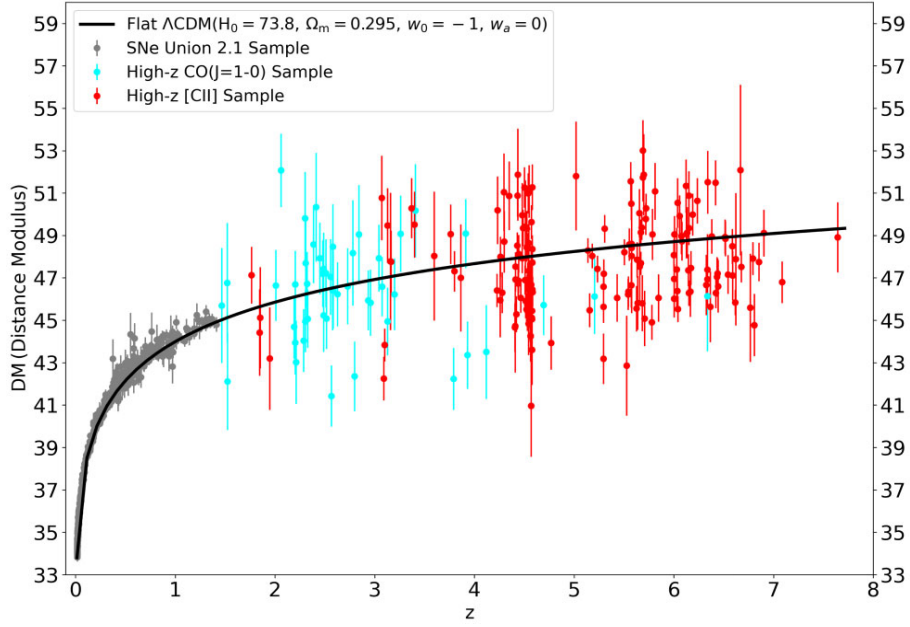

**Figure 4.** This diagram depicts the observed measured DM as a function of redshift $z$ (within $0.0 < z < 8.0$). The DM of high-$z$ [C II] sample, which was measured by the four steps in Section 5.1, is denoted in the red dot. The DM of the Union 2.1 sample is denoted in the grey dot. The DM of the high-$z$ CO ($J$ = 1-0) sample, which was provided in W19, is denoted in the cyan dot. For each data dot, the vertical bar represents the $1\sigma$ measurement uncertainty of the DM. The black curve demonstrates the spatially-flat concordance $\Lambda$CDM model derived in a set of the parameter values (i.e. $H_0 = 73.8 \text{ km s}^{-1} \text{ Mpc}^{-1}$, $\Omega_m = 0.295$, $w_0 = -1$, and $w_a = 0$).

There is a simple formula for the $w(z)$ according to the Chevallier–Polarski–Linder model (i.e. the CPL model; Chevallier & Polarski 2001; Linder 2003). Due to this CPL model, the $w(z)$ can be expressed as

$$w(z) = w_0 + w_a \times \left(\frac{z}{1+z}\right), \quad (31)$$

where the $w_0$ is a present-day parameter of the EoS of dark energy and the $w_a$ is an evolutionary coefficient to scale the fractional function of $z$. By substituting equation (31) into equation (30), one can specialize equation (29) into a form as

$$E(z|\Omega_m, \Omega_k, \Omega_{DE}, w_0, w_a)$$
$$= \sqrt{\Omega_m(1+z)^3 + \Omega_k(1+z)^2 + \Omega_{DE}(1+z)^{3(1+w_0+w_a)} \exp(\frac{-3w_a z}{1+z})}. \quad (32)$$

Herein, there are five free parameters (i.e. $\Omega_m$, $\Omega_k$, $\Omega_{DE}$, $w_0$, and $w_a$) in equation (32).

We simplified equation (32) by the two methods: (1) We still adopted the cosmology parameters used in Section 3.1 since both the values of the $H_0$ and $\Omega_m$ has been well-constrained and reliable, and (2) we assigned the $\Omega_k = 0$ and $\Omega_m + \Omega_{DE} = 1$ by referring to Suzuki et al. (2012). Because of these two methods, equation (32) were simplified into as a two-free-parameter function so that we were able to concentrate on constraining the $w_0$ and $w_a$ in the CPL model. Next, we performed the parameter constraint by using the $\chi^2$ approach. The approach is defined as

$$\chi^2 = \sum_{i=1}^{N} \left(\frac{\text{DM}_i^{\text{obs.}}(z_i) - \text{DM}_i^{\text{th.}}(z_i)}{\sigma_{\text{DM}_i^{\text{obs.}}}}\right)^2. \quad (33)$$

Here, the $N$ denotes the number of the sample involved in the $\chi^2$ summation, the $\text{DM}_i^{\text{obs.}}(z_i)$ denotes the $i$-th observed DM at the $i$-th redshift $z$, $\sigma_{\text{DM}_i^{\text{obs.}}}$ denotes the $i$-th observational error, and the $\text{DM}_i^{\text{th.}}(z_i)$ denotes the $i$-th theoretical DM at the $i$-th redshift. A likelihood function $L$ is also defined to relate the $\chi^2$ by

$$L \sim e^{(\frac{-\chi^2}{2})}. \quad (34)$$

Therefore, with the definition of equation (34) used in the parameter constraint algorithm, a maximum value of the $L$ function occurs while both the values of the $w_0$ and $w_a$ best fit to the sample data.

Herein, in the parameter constraint algorithm, we allocated three data sets: SNe (SNe data alone), SNe/CO [SNe + high-$z$ CO(1-0)], and SN/CO/CII [SNe + high-$z$ CO(1-0) + high-$z$ [C II]], and these three data sets were involve in the $\chi^2$ (the $L$) algorithm separately so as to examine the change of the constraint result due to the different involvement of the high-$z$ samples (i.e. the high-$z$ CO sample and high-$z$ [C II] sample).

Fig. 5 presents the constraint results for the SNe, SNe/CO, and SNe/CO/CII data sets. In the left-hand panel of Fig. 5, obviously, the confident-level contour shrinks as the size of the data set enlarges, indicating that the involvement of the measured DMs of the high-$z$ samples may help to reduce the constraint uncertainties of the $w_0$ and $w_a$. Especially, the contour shrinks significantly as the involvement of the high-$z$ [C II] sample.

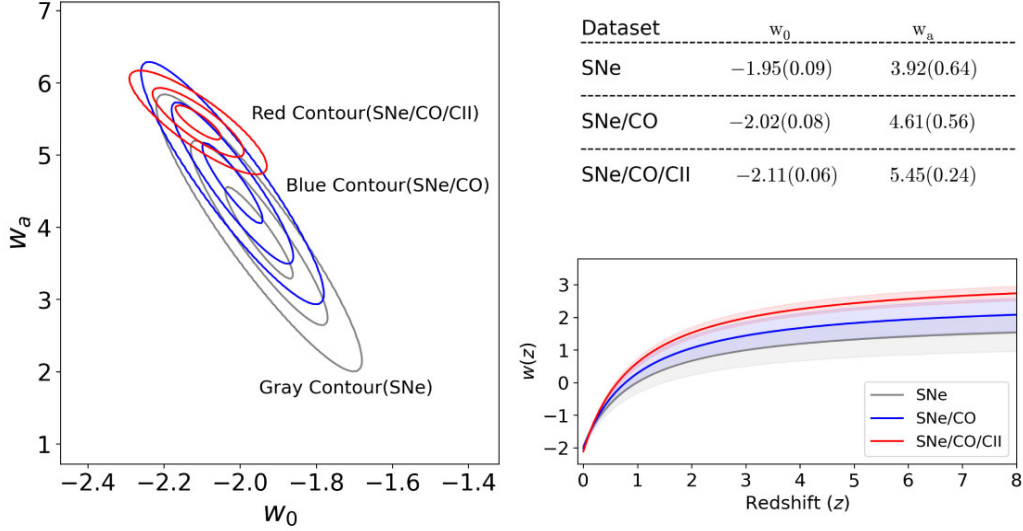

**Figure 5.** The left-hand panel illustrates the confidence-level contours (the inner, middle, and outer contours represent the $1\sigma$, $2\sigma$, and $3\sigma$ confidence levels) for the SNe (the grey contours), SNe/CO (the blue contours), and SNe/CO/CII (the red contours) data sets in the $w_0$–$w_a$ plane. The top-right table lists all the best-fitting values of the $w_0$ and $w_a$, with their $1\sigma$ uncertainties listed in the parentheses. Based on the constraint results in the top-right table, the bottom-right panel presents the evolutions of the $w(z)$ for the SNe, SNe/CO, and SNe/CO/CII.

We also depicted those constraint results presented in Fig. 5 in a normalized DM–$z$ diagram as shown in Fig. 6. In Fig. 6, the DMs of the sample data are normalized to these expected for the $\Lambda$CDM model as a form of $\frac{\text{DM}-\text{DM}_{\Lambda\text{CDM}}}{\text{DM}_{\Lambda\text{CDM}}}$. The orange, yellow, and green curves manifest the best-fitting curves (the possible cosmic spatial growing trends affect by the time-varying dark energy) for the SNe, SNe/CO, and SNe/CO/CII. The black horizontal line depicted along the $x$-axis manifests the $\Lambda$CDM model which is used to as the base line. After checking the three best-fitting curves in $z > 1$, we found the best-fitting curve of the SNe deviates from the $\Lambda$CDM base in a smallest degree and the one of the SNe/CO/CII in a largest degree, and the best-fitting curve of the SNe/CO places between the other two best-fitting curves. In other words, more the sample data are involved in the parameter constraint, more the deviations are manifested with respect to the black base line. It also indicates the high-$z$ [C II]-detected galaxies may be used to track the possible cosmic growing trend in $z > 1$ which distinguishes from the conventional cosmological model (i.e. the $\Lambda$CDM model). Even, like the proposal of tracking the time-varying dark energy by using gamma-ray bursts (e.g. Schaefer 2003; Firmani et al. 2006; Liang et al. 2008; Izzo et al. 2009; Tsutsui et al. 2009) or fast-radio bursts (e.g. Zhou et al. 2014; Walters et al. 2018; Jaroszynski 2019; Hashimoto et al. 2019, 2020), the high-$z$ [C II]-detected galaxies may be feasibly utilized to track the time-varying nature of dark energy.

## 6 CONCLUSION AND FUTURE WORK

In Section 4, the [C II] LFR was confirmed with a high significance, and its slope presented no redshift evolution. However, the intercept of the [C II] LFR shifted while the redshift range of the sample galaxy changed. Moreover, the data scatter about the [C II] LFR corresponded to that about the CO(1-0) LFR. Further, as demonstrated in Section 5, we realized that the [C II] LFR could be functioned as a new distance ladder on measuring the DMs to the galaxies in $z > 1$, and even those measured DMs could be possibly utilized for constraining the cosmological parameters. In conclusion, by our investigation, the [C II] LFR can be another robust scaling relation possibly useful in the astronomical measurement.

However, in Section 5, one should notice that the uncertainties of those measured DMs have a great influence on the accuracies of the constrained cosmological parameters. Due to it, the uncertainty of the measured DM could be sourced from three aspects which go around the statistical properties of the [C II] LFR, and we list them in the below and discuss.

### 6.1.1 *The uncertainties of the slope and intercept*

Through equations (19), (22), (24), and (26), the uncertainties of the slope (0.054) and intercept (0.023) of the calibrated [C II] LFR (see equation 17) were together transformed to the uncertainty of the DM. By tracing back in Section 5.1, equations (13) and (14) are a source to induce the slope and intercept uncertainties in equation (17), and it may be relevant to the way of using the distance calibrator (the Union 2.1 sample used in this paper). In Section 5.1, The best-fitting DM–$\log_{10}(z)$ relation line of equation (13) was singly obtained through fitting with the whole Union 2.1 sample data for the subsequent distance calibration. However, after checking the bottom-right panel and the central plot in Fig. 3, the number distribution of the Union 2.1 sample is not homogeneous in $0.015 < z < 1.414$ and the data scatter about that black best-fitting DM–$\log_{10}(z)$ relation becomes larger at higher redshifts [$\log_{10}(z) > -0.5$ or $z \gtrsim 0.31$]. These two factors may induce some potential uncertainties; however, they might be implicit while the single DM–$\log_{10}(z)$ relation was considered at that moment.

### 6.1.2 *The intercept discrepancy*

By using equation (1) in the [C II] observation, taking logarithm in the both sides of this equation, and transforming into a form similar to equation (12) with adding a pair of the $A$ and $B$ as the pivot point,

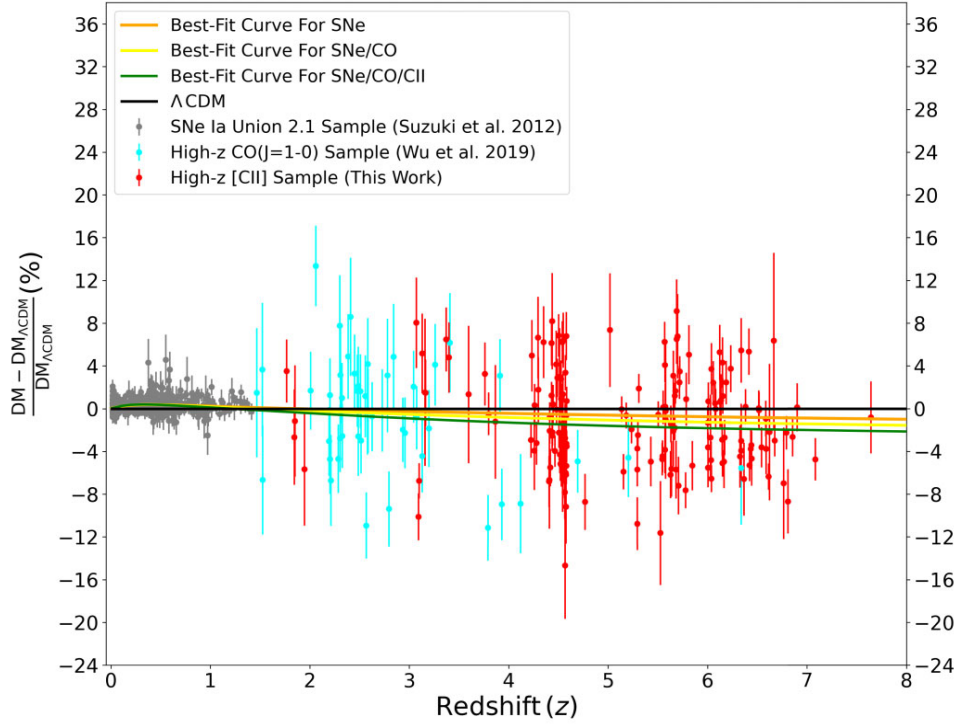

**Figure 6.** The results in Fig. 5 are depicted in this normalized DM–$z$ diagram. The DM is normalized to that expected by the ΛCDM model ($H_0 = 73.8 \, \mathrm{km \, s^{-1} \, Mpc^{-1}}$, $\Omega_\mathrm{m} = 0.295$, $w_0 = -1$, and $w_a = 0$) as a form of $\frac{\mathrm{DM} - \mathrm{DM}_{\Lambda\mathrm{CDM}}}{\mathrm{DM}_{\Lambda\mathrm{CDM}}}$. The orange, yellow, and green curves represent the normalized best-fitting DM($z$) curves (i.e. the possible cosmic spatial growing trends affect by the time-varying dark energy) for the SNe, SNe/CO, and SNe/CO/CII, respectively. The black horizontal line depicted along the $x$-axis represents the ΛCDM model.

one can obtain an identity as

$$\log_{10}(\mathrm{FWHM}_{[\mathrm{C\,II}]}) - \mathrm{A} = \left(\frac{1}{n}\right) \times \log_{10}\left(\frac{C}{f_\mathrm{gas}}\right) + \left(\frac{m}{n}\right) \times \left(\log_{10}(L'_{[\mathrm{C\,II}]}) - \mathrm{B}\right). \quad (35)$$

The intercept part of equation (35) may be evolved in redshift due to the definition of $f_\mathrm{gas} = \frac{M_\mathrm{mol}}{M_\mathrm{dyn}}$, and here the $M_\mathrm{dyn}$ is dependent on the galaxy star-formation rate which is evolves in redshift observationally. Moreover, the $M_\mathrm{dyn}$ is also relevant to the geometrical factor (denoted as $f$, Gnerucci et al. 2011) which denotes the galaxy dynamical structure that may differ in different redshifts. Recalling the data selection in Section 2, our [C II] sample galaxies involved in the analysis spanned a large range of redshift ($0.02 < z < 7.70$), and in actual their star-formation rates might widely extend in redshift and their dynamical structures were also expected to be sophisticated. By this, it may be the reason of the intercept shift (the intercept discrepancy) we found in Section 4. Besides, the existence of the intercept indicates that the [C II] LFR in $z < 1$ is different from the one in $z > 1$ although their slopes may be consistent closely, and this discrepancy could be an implicit uncertainty to cause the uncertainty of the DM while one applies the calibrated [C II] LFR in $z < 1$ to the high-redshift Universe.

*6.1.3 The assignment of the pivot point*

By checking equation (20) used in equation (19), one may notice that a smaller uncertainty of the DM is derived as long as the strength (the absolute value) of the $\rho$ becomes reduced. Moreover, in the third point listed in the end of Section 4, a smaller absolute value of the $\rho$ was found for the case of using their own pivot points. With the hint in equations (20) and (19) and that finding in our LR analysis, it is an implication that the way of assigning the pivot point in the LR model may be a technique to reduce the DM uncertainty while the procedure demonstrated in Section 5.1 is applied.

The three aspects discussed previously will boost us to come up with some new methods which are different from the ones adopted in this paper if we are going to further constrain the cosmological parameters in a high-accuracy level for comparing with the recent published result (e.g. the cosmic microwave background observation in *Planck*; Planck Collaboration VI 2018). For instance, we may consider more criteria on diving [C II] sample (i.e. specific ranges of the redshift, IR luminosity, star-formation rate, and geometric factor) so as to check the statistical behaviours of the slope and intercept deeply. And, we may derive multiple best-fitting DM–$\log_{10}(z)$ relations in different redshift ranges in order to lower the potential uncertainties due to the use of the single DM–$\log_{10}(z)$ relation in the distance calibration. Moreover, with the on-going survey project (e.g. the REBELS project in *ALMA*) and the new-generation telescope (e.g. *JWST*), more new galaxies locating in different cosmic redshifts will be observed, and their observed data are also a help for our further investigation about the property of the [C II] LFR.

**ACKNOWLEDGEMENTS**

We greatly appreciate the anonymous referee for the careful reading and helpful suggestions that significantly improved this paper. During the review of this work, we were deeply saddened that Prof. Yu Gao passed away in 2022 May due to a sudden illness. YHW and JFW would like to dedicate this paper to the memory of Prof. Yu Gao for his kind support and mentoring. YHW acknowledges the collaborative study with Prof. T. Goto during the PhD thesis work, which further enlightened this work. We acknowledge the support of the National Natural Science Foundation of China (Grant numbers are 12033004, 12221003, U1831205, 11861131007, and U1731237) and the National Key Research and Development Program (Grant No. 2017YFA0402704). YHW acknowledges the support of the 2021 International Postdoctoral Exchange Fellowship Program (Talent Introduction Program, Grant No. YJ20210131).

## DATA AVAILABILITY

The data that were used in the linear regression analysis had been described in Section 2. Readers can also check the data used for the linear regression analysis in Appendix of this paper.

# APPENDIX A: SAMPLE GALAXIES AND AVAILABLE DATA

**Table A1.** The observed data of the 24 ULIRGs selected from F13. Column 1: The identification (ID) number of the ULIRG; Column 2: The redshift published in the reference; Column 3: The [C II] flux in the unit of Jy km s$^{-1}$; Column 4: The [C II] flux in the unit of $\times 10^{-21}$ W cm$^{-2}$; Column 5: The [C II] linewidth in km s$^{-1}$. The flux values in Column 3 are converted from these in Column 4 by using equation (3). The values located behind the $\pm$ signs represent the 1$\sigma$ measurement uncertainties which are available in F13.

| ULIRG ID | $z^a$ | Flux$_{[C II]}$ (Jy km s$^{-1}$) | Flux$_{[C II]}$ ($10^{-21}$ W cm$^{-2}$) | Linewidth$_{[C II]}$ (km s$^{-1}$) |
| --- | --- | --- | --- | --- |
| Mrk 463 | 0.051 | 3365 $\pm$ 139 | 21.33 $\pm$ 0.88 | 250 $\pm$ 2 |
| IRAS 19254−7245 | 0.063 | 4948 $\pm$ 303 | 31.37 $\pm$ 1.92 | 514 $\pm$ 12 |
| IRAS 06035−7102 | 0.079 | 4824 $\pm$ 322 | 30.58 $\pm$ 2.04 | 277 $\pm$ 2 |
| IRAS 20414−1651 | 0.087 | 1440 $\pm$ 73 | 9.13 $\pm$ 0.46 | 475 $\pm$ 12 |
| IRAS 06206−6315 | 0.092 | 1459 $\pm$ 188 | 9.25 $\pm$ 1.19 | 369 $\pm$ 10 |
| IRAS 08311−2459 | 0.100 | 3836 $\pm$ 55 | 24.32 $\pm$ 0.35 | 272 $\pm$ 3 |
| IRAS 15462−0450 | 0.100 | 1167 $\pm$ 125 | 7.40 $\pm$ 0.79 | 163 $\pm$ 4 |
| IRAS 20087−0308 | 0.106 | 2704 $\pm$ 28 | 17.14 $\pm$ 0.18 | 550 $\pm$ 5 |
| IRAS 11095−0238 | 0.107 | 1185 $\pm$ 186 | 7.51 $\pm$ 1.18 | 296 $\pm$ 9 |
| IRAS 23230−6926 | 0.107 | 1576 $\pm$ 32 | 9.99 $\pm$ 0.20 | 307 $\pm$ 5 |
| IRAS 01003−2238 | 0.118 | 845 $\pm$ 194 | 5.36 $\pm$ 1.23 | 150 $\pm$ 7 |
| IRAS 13451+1232 | 0.121 | 707 $\pm$ 33 | 4.48 $\pm$ 0.21 | 465 $\pm$ 18 |
| IRAS 00188−0856 | 0.128 | 607 $\pm$ 161 | 3.85 $\pm$ 1.02 | 279 $\pm$ 13 |
| IRAS 12071−0444 | 0.128 | 1022 $\pm$ 142 | 6.48 $\pm$ 0.9 | 250 $\pm$ 6 |
| IRAS 20100−4156 | 0.130 | 1522 $\pm$ 30 | 9.65 $\pm$ 0.19 | 247 $\pm$ 4 |
| IRAS 23253−5415 | 0.130 | 1983 $\pm$ 170 | 12.57 $\pm$ 1.08 | 473 $\pm$ 8 |
| IRAS 03158+4227 | 0.134 | 888 $\pm$ 17 | 5.63 $\pm$ 0.11 | 243 $\pm$ 4 |
| IRAS 16090−0139 | 0.134 | 1505 $\pm$ 189 | 9.54 $\pm$ 1.20 | 352 $\pm$ 8 |
| IRAS 10378+1109 | 0.136 | 557 $\pm$ 32 | 3.53 $\pm$ 0.2 | 409 $\pm$ 20 |
| IRAS 07598+6508 | 0.148 | 331 $\pm$ 60 | 2.10 $\pm$ 0.38 | 197 $\pm$ 35 |
| IRAS 03521+0028 | 0.152 | 592 $\pm$ 170 | 3.75 $\pm$ 1.08 | 442 $\pm$ 27 |
| 3C 273 | 0.158 | 232 $\pm$ 38 | 1.47 $\pm$ 0.24 | 465 $\pm$ 66 |
| Mrk 1014 | 0.163 | 683 $\pm$ 169 | 4.33 $\pm$ 1.07 | 316 $\pm$ 17 |
| IRAS 00397−1312 | 0.262 | 254 $\pm$ 39 | 1.61 $\pm$ 0.25 | 427 $\pm$ 61 |

*Note.*
[a] Owing to unavailable redshift uncertainty in the referenced paper, we assign zero to the redshift uncertainty.

**Table A2.** The observed data of the seven ULIRGs selected from M14. Column 1: The ID of the ULIRG; Column 2: The redshift published in the reference; Column 3: The [C II] flux in the unit of Jy km s$^{-1}$; Column 4: The [C II] flux in the unit of $\times 10^{-21}$ W m$^{-2}$; Column 5: The [C II] FWHM linewidth in the unit of km s$^{-1}$. The flux values in Column 3 are converted from these in Column 4 by using equation (3). The values located behind the $\pm$ signs represent the 1$\sigma$ measurement uncertainties which are available in M14.

| Object ID | $z^a$ | Flux$_{[C II]}$ (Jy km s$^{-1}$) | Flux$_{[C II]}$ ($10^{-21}$ W m$^{-2}$) | FWHM$_{[C II]}$ (km s$^{-1}$) |
| --- | --- | --- | --- | --- |
| BOOTES3 | 0.219 | 752.4 $\pm$ 88.3 | 47.7 $\pm$ 5.6 | 545 $\pm$ 43 |
| CDFS2 | 0.248 | 323.4 $\pm$ 44.2 | 20.5 $\pm$ 2.8 | 390 $\pm$ 54 |
| BOOTES2 | 0.250 | 197.2 $\pm$ 55.2 | 12.5 $\pm$ 3.5 | 661 $\pm$ 149 |
| SWIRE4 | 0.251 | 381.7 $\pm$ 55.2 | 24.2 $\pm$ 3.5 | 556 $\pm$ 75 |
| ELAISS | 0.265 | 348.6 $\pm$ 33.1 | 22.1 $\pm$ 2.1 | 433 $\pm$ 41 |
| CDFS1 | 0.289 | 263.4 $\pm$ 45.7 | 16.7 $\pm$ 2.9 | 764 $\pm$ 132 |
| SWIRE5 | 0.366 | 438.5 $\pm$ 33.1 | 27.8 $\pm$ 2.1 | 417 $\pm$ 43 |

*Note.*
[a] Owing to unavailable redshift uncertainty in the referenced paper, we assign zero to the redshift uncertainty.

**Table A3.** The observed data of the 26 [C II]-detected galaxies selected from I15. Column 1: The ID of the galaxy; Column 2: The redshift published in the reference; Column 3: The [C II] flux in the unit of Jy km s$^{-1}$; Column 4: The [C II] FWHM linewidth in the unit of km s$^{-1}$. The values located behind the ± signs represent the 1$\sigma$ measurement uncertainties which are available in I15.

| Galaxy ID | $z^a$ | Flux$_{[CII]}$ (Jy km s$^{-1}$) | FWHM$_{[CII]}$ (km s$^{-1}$) |
|---|---|---|---|
| G09.v2.52 | 0.026 | 1331 ± 33 | 168 ± 11.33 |
| G09.v2.235 | 0.027 | 901 ± 34 | 142 ± 19.93 |
| G09.v2.66 | 0.031 | 2090 ± 35 | 177 ± 8.13 |
| G09.v2.23 | 0.033 | 2373 ± 44 | 356 ± 8.34 |
| G09.v2.103 | 0.041 | 1509 ± 31 | 220 ± 0.1 |
| G09.v2.137 | 0.044 | 2349 ± 42 | 284 ± 8.39 |
| G09.v2.170 | 0.051 | 1439 ± 38 | 195 ± 12.97 |
| G09.v2.45 | 0.051 | 2274 ± 41 | 266 ± 7.28 |
| G09.v2.58 | 0.052 | 1832 ± 39 | 223 ± 9.54 |
| G09.v2.80 | 0.053 | 907 ± 32 | 104 ± 15.06 |
| G09.v2.117 | 0.054 | 898 ± 31 | 168 ± 12.24 |
| G09.v2.55 | 0.054 | 1697 ± 53 | 296 ± 15.46 |
| G09.v2.42 | 0.055 | 1291 ± 43 | 215 ± 17.74 |
| G09.v2.38 | 0.059 | 1794 ± 58 | 527 ± 16.39 |
| G09.v2.60 | 0.060 | 1997 ± 55 | 482 ± 13.53 |
| G09.v2.175 | 0.070 | 1235 ± 46 | 312 ± 17.73 |
| G09.v2.102 | 0.073 | 849 ± 46 | 356 ± 22.88 |
| G09.v2.299 | 0.074 | 880 ± 48 | 258 ± 25.34 |
| G09.v2.111 | 0.078 | 1052 ± 42 | 192 ± 18.32 |
| G09.v2.77 | 0.079 | 1917 ± 58 | 403 ± 14.68 |
| G09.v2.232 | 0.096 | 648 ± 32 | 184 ± 0.01 |
| G09.v2.90 | 0.133 | 909 ± 91 | 459 ± 55.95 |
| G09.v2.76 | 0.107 | 1201 ± 47 | 230 ± 18.52 |
| G09.v2.107 | 0.128 | 1467 ± 62 | 227 ± 21.31 |
| G09.v2.48 | 0.072 | 2398 ± 43 | 223 ± 7.67 |
| G09.v2.26 | 0.182 | 1052 ± 51 | 103 ± 23.42 |

*Note.*
$^a$Owing to unavailable redshift uncertainty in the referenced paper, we assign zero to the redshift uncertainty.

**Table A4.** The observed data of the four [C II]-detected galaxies selected from Z18. Column 1: The ID of the galaxy; Column 2: The spectroscopic redshift measured in the [C II] emission line; Column 3: The [C II] flux in the unit of Jy km s$^{-1}$; Column 4: The [C II] FWHM linewidth in the unit of km s$^{-1}$; Column 5: The flux density in the unit of [C II], mJy; Column 6: The [C II] luminosity in the unit of $10^9$ L$_\odot$; Column 7: The luminosity distance (in Mpc) based on ($\Omega_m$, $\Omega_\Lambda$, and H$_0$) = (0.3, 0.7, and 70). By using equations (6) and (7) in Section 3.2, the result in Column 3 is derived from the results in Columns 6 and 7, and the result in Column 4 is derived from the results in Columns 3 and 5 by using equations (9) and (10). In Columns 5 and 6, the uncertainty values located behind the ± signs are available in Z18.

| Galaxy ID | $z$ | $I_{[CII]}$ (Jy km s$^{-1}$) | FWHM (km s$^{-1}$) | $F$ ([C II], mJy) | $L_{[CII]}$ ($10^9$ L$_\odot$) | Distance (Mpc) |
|---|---|---|---|---|---|---|
| 9347 | 1.8505 ± 0.0002 | 7.225 ± 2.282 | 319.106 ± 142.617 | 21.28 ± 6.73 | 0.95 ± 0.3 | 14 128.014 |
| 6515 | 1.8438 ± 0.0002 | 7.273 ± 1.951 | 278.988 ± 105.829 | 24.50 ± 6.57 | 1.23 ± 0.33 | 14 065.253 |
| 10076 | 1.9462 ± 0.0006 | 6.599 ± 2.090 | 213.644 ± 95.403 | 29.03 ± 9.14 | 2.40 ± 0.76 | 15 029.203 |
| 9834 | 1.7644 ± 0.0003 | 7.877 ± 1.160 | 482.607 ± 99.432 | 15.34 ± 2.21 | 1.29 ± 0.19 | 13 324.937 |

**Table A5.** The observed data of the 23 quasars selected from D18. Column 1: The ID of the quasar; Column 2: The spectroscopic redshift measured in the [C II] emission; Column 3: The [C II] flux in the unit of Jy km s$^{-1}$; Column 4: The [C II] FWHM linewidth in the unit of km s$^{-1}$. The values located behind the ± signs represent the 1$\sigma$ measurement uncertainties which are available in D18.

| Quasar ID | $z$ | Flux$_{[C II]}$ (Jy km s$^{-1}$) | FWHM (km s$^{-1}$) |
|---|---|---|---|
| PJ007+04 | 6.0008 ± 0.0004 | 1.58 ± 0.08 | 340 ± 36 |
| PJ009−10 | 6.0039 ± 0.0004 | 2.49 ± 0.15 | 251 ± 34 |
| J1306+0356 | 6.0337 ± 0.0004 | 1.63 ± 0.09 | 246 ± 31 |
| J1207+0630 | 6.0366 ± 0.0009 | 0.92 ± 0.12 | 489 ± 78 |
| J0454−4448 | 6.0581 ± 0.0006 | 0.85 ± 0.08 | 426 ± 57 |
| J0842+1218 | 6.0763 ± 0.0005 | 1.44 ± 0.11 | 396 ± 45 |
| J2100−1715 | 6.0812 ± 0.0005 | 1.52 ± 0.14 | 382 ± 51 |
| J1509−1749 | 6.1225 ± 0.0007 | 1.50 ± 0.12 | 631 ± 72 |
| PJ065−19 | 6.1247 ± 0.0006 | 0.69 ± 0.08 | 345 ± 67 |
| J2318−3029 | 6.1458 ± 0.0004 | 2.34 ± 0.12 | 320 ± 34 |
| PJ217−16 | 6.1498 ± 0.0011 | 0.70 ± 0.11 | 491 ± 75 |
| PJ359−06 | 6.1722 ± 0.0004 | 2.47 ± 0.16 | 330 ± 39 |
| PJ065−26 | 6.1877 ± 0.0005 | 2.05 ± 0.11 | 517 ± 44 |
| PJ308−21 | 6.2341 ± 0.0005 | 1.79 ± 0.1 | 570 ± 45 |
| J0142−3327 | 6.3379 ± 0.0004 | 2.62 ± 0.06 | 300 ± 32 |
| J2211−3206 | 6.3394 ± 0.001 | 0.57 ± 0.11 | 529 ± 118 |
| J1152+0055 | 6.3643 ± 0.0005 | 0.54 ± 0.07 | 167 ± 45 |
| PJ159−02 | 6.3809 ± 0.0005 | 1.15 ± 0.07 | 373 ± 40 |
| PJ183+05 | 6.4386 ± 0.0004 | 5.84 ± 0.08 | 374 ± 30 |
| J2318−3113 | 6.4435 ± 0.0005 | 1.11 ± 0.14 | 234 ± 49 |
| PJ167−13 | 6.5148 ± 0.0005 | 2.53 ± 0.07 | 437 ± 34 |
| PJ231−20 | 6.5864 ± 0.0005 | 2.65 ± 0.12 | 404 ± 39 |
| J1048−0109 | 6.6759 ± 0.0005 | 2.52 ± 0.07 | 330 ± 33 |

**Table A6.** The observed data of the 30 [C II]-detected galaxies selected from C20. Column 1: The ID of the selected galaxy; Column 2: The redshift published in the reference. The redshift uncertainty is referred to Strandet et al. (2016); Column 3: The [C II] flux in the unit of Jy km s$^{-1}$; Column 4: The [C II] FWHM linewidth in the unit of km s$^{-1}$; Column 5: The magnification factor $\mu$ of the selected galaxy due to the gravitational-lens effect. The values located behind the ± signs represent the 1$\sigma$ measurement uncertainties which are available in C20.

| Object ID | $z$ | Flux$_{[C II]}$ (Jy km s$^{-1}$) | FWHM$_{[C II]}$ (km s$^{-1}$) | $\mu$ |
| --- | --- | --- | --- | --- |
| SPT2357−51 | 3.0703 ± 0.0006 | 9.1 ± 4.6 | 743 ± 202 | 2.9 ± 0.1 |
| SPT0103−45 | 3.0917 ± 0.0003 | 190.4 ± 23.8 | 239 ± 39 | 5.1 ± 0.1 |
| SPT0550−53 | 3.128 ± 0.0007 | 88.1 ± 8.6 | 789 ± 165 | 6.3 ± 1.0 |
| SPT2101−60 | 3.1560[a] | 9.3 ± 13.8 | 353 ± 189 | 6.3 ± 1.0 |
| SPT0551−50 | 3.164 ± 0.001 | 216.1 ± 16.3 | 734 ± 95 | 4.5 ± 0.5 |
| SPT0529−54 | 3.3689 ± 0.0001 | 64.6 ± 7.7 | 733 ± 81 | 13.2 ± 0.8 |
| SPT0532−50 | 3.3988 ± 0.0001 | 91.7 ± 11.2 | 719 ± 125 | 10.0 ± 0.6 |
| SPT0300−46 | 3.5954 ± 0.0007 | 15.7 ± 5.1 | 414 ± 237 | 5.7 ± 0.4 |
| SPT2147−50 | 3.7602 ± 0.0003 | 24.4 ± 7.5 | 534 ± 104 | 6.6 ± 0.4 |
| SPT2340−59 | 3.864 ± 0.001 | 48.1 ± 8.5 | 473 ± 220 | 3.4 ± 0.3 |
| SPT0418−47 | 4.2248 ± 0.0007 | 138.1 ± 10.4 | 322 ± 37 | 32.7 ± 0.7 |
| SPT0113−46 | 4.2328 ± 0.0005 | 49.7 ± 12.8 | 578 ± 136 | 23.9 ± 0.5 |
| SPT2052−56 | 4.257[a] | 14.6 ± 1.8 | 382 ± 122 | 1.0 ± 0.0 |
| SPT2311−54 | 4.2795 ± 0.0004 | 45.3 ± 4.6 | 352 ± 52 | 6.3 ± 1.0 |
| SPT0345−47 | 4.2958 ± 0.0002 | 15.4 ± 4.3 | 669 ± 177 | 8.0 ± 0.5 |
| SPT0136−63 | 4.299[a] | 33.3 ± 2.9 | 526 ± 115 | 6.3 ± 1.0 |
| SPT0155−62 | 4.349[a] | 26.3 ± 6.2 | 760 ± 127 | 6.3 ± 1.0 |
| SPT2103−60 | 4.4357 ± 0.0006 | 15.6 ± 10.4 | 602 ± 204 | 27.8 ± 1.8 |
| SPT0441−46 | 4.4771 ± 0.0006 | 26.3 ± 5.8 | 546 ± 123 | 12.7 ± 1.0 |
| SPT0319−47 | 4.510 ± 0.004 | 11.4 ± 10.5 | 562 ± 182 | 2.9 ± 0.3 |
| SPT2146−55 | 4.5672 ± 0.0002 | 11.3 ± 5.8 | 277 ± 73 | 6.6 ± 0.4 |
| SPT2132−58 | 4.7677 ± 0.0002 | 35.9 ± 6.9 | 212 ± 43 | 5.7 ± 0.5 |
| SPT0202−61 | 5.018[a] | 19.3 ± 4.1 | 771 ± 325 | 9.1 ± 0.07 |
| SPT2319−55 | 5.2929 ± 0.0005 | 39.2 ± 4.7 | 176 ± 28 | 6.9 ± 0.6 |
| SPT2353−50 | 5.576 ± 0.003 | 20.7 ± 6.4 | 429 ± 247 | 6.3 ± 1.0 |
| SPT0245−63 | 5.626[a] | 26.9 ± 4.5 | 383 ± 65 | 1.0 ± 0.01 |
| SPT0346−52 | 5.6559[a] | 64.1 ± 8.2 | 486 ± 85 | 5.6 ± 0.1 |
| SPT0348−62 | 5.656[a] | 19.6 ± 3.4 | 506 ± 132 | 1.2 ± 0.01 |
| SPT0243−49 | 5.699 ± 0.001 | 17.4 ± 2.7 | 796 ± 202 | 6.7 ± 0.5 |
| SPT2351−57 | 5.811 ± 0.002 | 5.4 ± 2.7 | 539 ± 82 | 6.3 ± 1.0 |

*Note.*
[a]Owing to unavailable redshift uncertainty in the referenced paper, we assign zero to the redshift uncertainty.

**Table A7.** The observed data of the 75 [C II]-detected galaxies in the ALPINE data set (https://caltech.box.com/s/eo4v0zaphnbf3t0v1761pcos5gofmnx6, Béthermin et al. 2020; Faisst et al. 2020; Le Fèvre et al. 2020; ). Column 1: The ID of the selected galaxy; Column 2: The redshift listed in the ALPINE data set; Column 3: The [C II] flux in the unit of Jy km s$^{-1}$; Column 4: The [C II] FWHM linewidth in the unit of km s$^{-1}$. The values located behind the ± signs represent the 1$\sigma$ measurement uncertainties which are available in the ALPINE data set.

| Galaxy ID | $z$ | Flux$_{[C II]}$ (Jy km s$^{-1}$) | FWHM$_{[C II]}$ (km s$^{-1}$) |
| --- | --- | --- | --- |
| DEIMOS COSMOS 432340 | 4.405[a] | 0.604 ± 0.145 | 151 ± 13 |
| CANDELS GOODSS 32 | 4.411[a] | 1.382 ± 0.136 | 279 ± 13 |
| DEIMOS COSMOS 709575 | 4.412[a] | 0.504 ± 0.109 | 254 ± 16 |
| vuds efdcs 530029038 | 4.430[a] | 1.169 ± 0.131 | 367 ± 14 |
| CANDELS GOODSS 12 | 4.431 ± 0.001 | 0.840 ± 0.251 | 541 ± 18 |
| DEIMOS COSMOS 422677 | 4.438[a] | 0.705 ± 0.117 | 233 ± 14 |
| DEIMOS COSMOS 630594 | 4.440[a] | 1.040 ± 0.104 | 260 ± 13 |
| vuds cosmos 510786441 | 4.464[a] | 1.096 ± 0.117 | 224 ± 13 |
| DEIMOS COSMOS 274035 | 4.479[a] | 0.444 ± 0.121 | 266 ± 19 |
| DEIMOS COSMOS 434239 | 4.488[a] | 2.314 ± 0.273 | 497 ± 13 |
| vuds cosmos 510605533 | 4.502 ± 0.001 | 0.426 ± 0.106 | 504 ± 31 |
| DEIMOS COSMOS 834764 | 4.506[a] | 0.921 ± 0.187 | 254 ± 17 |
| DEIMOS COSMOS 493583 | 4.513[a] | 0.703 ± 0.102 | 198 ± 13 |
| vuds cosmos 5100822662 | 4.521[a] | 1.278 ± 0.105 | 208 ± 13 |
| DEIMOS COSMOS 308643 | 4.525[a] | 0.924 ± 0.144 | 406 ± 40 |
| DEIMOS COSMOS 665509 | 4.526[a] | 1.375 ± 0.311 | 372 ± 17 |
| DEIMOS COSMOS 680104 | 4.530[a] | 0.893 ± 0.272 | 190 ± 16 |
| vuds cosmos 5180966608 | 4.530[a] | 2.226 ± 0.176 | 243 ± 13 |
| DEIMOS COSMOS 859732 | 4.532 ± 0.001 | 0.697 ± 0.165 | 326 ± 16 |
| DEIMOS COSMOS 628063 | 4.533 ± 0.001 | 0.187 ± 0.052 | 167 ± 26 |
| DEIMOS COSMOS 627939 | 4.534[a] | 1.173 ± 0.102 | 252 ± 13 |
| DEIMOS COSMOS 400160 | 4.540 ± 0.001 | 0.630 ± 0.162 | 518 ± 19 |
| DEIMOS COSMOS 880016 | 4.542[a] | 0.894 ± 0.120 | 274 ± 14 |
| DEIMOS COSMOS 803480 | 4.542[a] | 0.214 ± 0.063 | 125 ± 18 |
| DEIMOS COSMOS 396844 | 4.542[a] | 1.863 ± 0.165 | 287 ± 13 |
| DEIMOS COSMOS 733857 | 4.545[a] | 0.777 ± 0.124 | 226 ± 13 |
| DEIMOS COSMOS 873756 | 4.546[a] | 5.837 ± 0.210 | 526 ± 13 |
| DEIMOS COSMOS 510660 | 4.548 ± 0.001 | 0.708 ± 0.197 | 540 ± 79 |
| vuds cosmos 5100537582 | 4.550[a] | 0.714 ± 0.108 | 206 ± 15 |
| vuds cosmos 5110377875 | 4.551[a] | 2.766 ± 0.170 | 234 ± 13 |
| DEIMOS COSMOS 842313 | 4.554 ± 0.001 | 0.731 ± 0.130 | 250 ± 15 |
| DEIMOS COSMOS 403030 | 4.559[a] | 0.745 ± 0.156 | 168 ± 14 |
| DEIMOS COSMOS 818760 | 4.561[a] | 6.917 ± 0.272 | 276 ± 13 |
| vuds cosmos 5100559223 | 4.563[a] | 0.705 ± 0.126 | 143 ± 13 |
| vuds cosmos 5100541407 | 4.563[a] | 1.901 ± 0.190 | 177 ± 13 |
| vuds cosmos 510596653 | 4.568[a] | 0.277 ± 0.052 | 62 ± 13 |
| vuds cosmos 5101209780 | 4.570[a] | 1.170 ± 0.250 | 453 ± 13 |
| vuds cosmos 5101218326 | 4.574[a] | 2.923 ± 0.125 | 229 ± 13 |
| vuds cosmos 5101210235 | 4.576[a] | 0.358 ± 0.101 | 145 ± 13 |
| DEIMOS COSMOS 665626 | 4.577[a] | 0.256 ± 0.068 | 102 ± 13 |
| DEIMOS COSMOS 881725 | 4.578[a] | 1.094 ± 0.101 | 198 ± 13 |
| vuds cosmos 5100969402 | 4.579[a] | 0.832 ± 0.087 | 291 ± 14 |
| vuds cosmos 5100994794 | 4.580[a] | 0.889 ± 0.081 | 230 ± 13 |
| vuds cosmos 5101244930 | 4.580 ± 0.001 | 0.809 ± 0.174 | 577 ± 35 |
| DEIMOS COSMOS 814483 | 4.581[a] | 1.265 ± 0.293 | 360 ± 14 |
| DEIMOS COSMOS 454608 | 4.583[a] | 1.027 ± 0.172 | 232 ± 15 |
| DEIMOS COSMOS 372292 | 5.136 ± 0.001 | 0.526 ± 0.066 | 289 ± 14 |
| DEIMOS COSMOS 873321 | 5.154[a] | 1.266 ± 0.142 | 201 ± 13 |
| DEIMOS COSMOS 539609 | 5.182 ± 0.001 | 0.652 ± 0.085 | 287 ± 14 |
| DEIMOS COSMOS 494763 | 5.234[a] | 0.633 ± 0.073 | 253 ± 14 |
| DEIMOS COSMOS 848185 | 5.293[a] | 2.056 ± 0.122 | 275 ± 13 |
| DEIMOS COSMOS 845652 | 5.307 ± 0.001 | 0.448 ± 0.117 | 339 ± 17 |
| DEIMOS COSMOS 378903 | 5.430 ± 0.001 | 0.225 ± 0.061 | 155 ± 13 |
| DEIMOS COSMOS 552206 | 5.502 ± 0.001 | 1.836 ± 0.140 | 366 ± 13 |
| CANDELS GOODSS 42 | 5.525 ± 0.001 | 0.066 ± 0.021 | 64 ± 13 |
| DEIMOS COSMOS 683613 | 5.542 ± 0.001 | 0.954 ± 0.082 | 216 ± 13 |
| DEIMOS COSMOS 494057 | 5.545 ± 0.001 | 0.861 ± 0.056 | 217 ± 13 |
| CANDELS GOODSS 14 | 5.553 ± 0.001 | 0.147 ± 0.039 | 230 ± 16 |
| CANDELS GOODSS 75 | 5.567 ± 0.002 | 0.386 ± 0.094 | 503 ± 15 |
| CANDELS GOODSS 21 | 5.572 ± 0.001 | 0.189 ± 0.062 | 167 ± 13 |
| CANDELS GOODSS 38 | 5.572 ± 0.001 | 0.313 ± 0.087 | 392 ± 28 |
| CANDELS GOODSS 47 | 5.575 ± 0.001 | 0.165 ± 0.052 | 237 ± 38 |

Note.

[a]According to Béthermin et al. (2020), the typical value of the redshift uncertainty is 0.0005 in ALPINE. Here we assign 0.001 to the redshift uncertainty.

**Table A8.** The continuous table of Table A7.

| Galaxy ID | $z$ | Flux$_{[C\,II]}$ (Jy km s$^{-1}$) | FWHM$_{[C\,II]}$ (km s$^{-1}$) |
|---|---|---|---|
| DEIMOS COSMOS 519281 | 5.576 ± 0.001 | 0.636 ± 0.128 | 282 ± 14 |
| DEIMOS COSMOS 416105 | 5.631 ± 0.001 | 0.156 ± 0.035 | 202 ± 14 |
| DEIMOS COSMOS 742174 | 5.636 ± 0.001 | 0.173 ± 0.042 | 139 ± 14 |
| DEIMOS COSMOS 417567 | 5.670 ± 0.001 | 0.362 ± 0.063 | 310 ± 16 |
| DEIMOS COSMOS 488399 | 5.670 ± 0.001 | 1.243 ± 0.058 | 303 ± 13 |
| DEIMOS COSMOS 773957 | 5.677 ± 0.001 | 0.481 ± 0.064 | 344 ± 13 |
| DEIMOS COSMOS 357722 | 5.684 ± 0.001 | 0.148 ± 0.049 | 134 ± 13 |
| DEIMOS COSMOS 430951 | 5.688 ± 0.002 | 0.663 ± 0.173 | 745 ± 77 |
| DEIMOS COSMOS 536534 | 5.689 ± 0.001 | 0.902 ± 0.182 | 621 ± 21 |
| DEIMOS COSMOS 351640 | 5.706 ± 0.001 | 0.269 ± 0.058 | 132 ± 13 |
| DEIMOS COSMOS 722679 | 5.717 ± 0.001 | 0.279 ± 0.103 | 331 ± 23 |
| vuds cosmos 5101288969 | 5.721 ± 0.001 | 0.107 ± 0.037 | 298 ± 18 |
| DEIMOS COSMOS 843045 | 5.847 ± 0.001 | 0.264 ± 0.076 | 158 ± 14 |

**Table A9.** The observed data of the 37 [C II]-detected galaxies selected from different source references. Column 1: The ID of the selected galaxy; Column 2: The redshift published in the reference; Column 3: The [C II] flux in the unit of Jy km s$^{-1}$; Column 4: The [C II] FWHM linewidth in the unit of km s$^{-1}$; Column 5: The magnification factor $\mu$ of the selected galaxy due to the gravitational-lens effect, and we assigned $\mu = 1$ in case the selected galaxy presented no sign of gravitational magnification in the source reference; Column 6: The galaxy type; Column 7: The source reference. The values located behind the ± signs represent the 1$\sigma$ measurement uncertainties which are available in the source references.

| Galaxy ID | $z$ | Flux$_{[C\,II]}$ (Jy km s$^{-1}$) | FWHM (km s$^{-1}$) | $\mu$ | Type | Reference |
|---|---|---|---|---|---|---|
| LAB1-ALMA3 | 3.0993 ± 0.0004 | 16.8 ± 2.1 | 275 ± 30 | 1 | Star-forming galaxy | Umehata et al. (2017) |
| ALMAJ120110.26+211756.2 | 3.7978 ± 0.0001 | 1.9 ± 0.3 | 330 ± 50 | 1 | Star-forming galaxy | Neeleman et al. (2017) |
| ALMAJ081740.86+135138.2 | 4.2601 ± 0.0001 | 5.4 ± 0.6 | 460 ± 50 | 1 | Star-forming galaxy | Neeleman et al. (2017) |
| BRI1335−0417 | 4.4074[a] | 26.6 ± 4.3 | 340 ± 140 | 1 | QSO | Wagg et al. (2010b) |
| ALESS 61.1 | 4.4189 ± 0.0004 | 2.5 ± 0.4 | 230 ± 25 | 1 | SMG | Swinbank et al. (2012) |
| ALESS 65.1 | 4.4445 ± 0.0005 | 5.4 ± 0.7 | 470 ± 35 | 1 | SMG | Swinbank et al. (2012) |
| LBG-1 | 5.29359 ± 0.00015 | 2.1 ± 0.2 | 230 ± 20 | 1 | Star-forming galaxy | Pavesi et al. (2016) |
| AzTEC-3 | 5.29795 ± 0.00013 | 7.8 ± 0.4 | 410 ± 15 | 1 | Star-forming galaxy | Pavesi et al. (2016) |
| HZ10 | 5.6543 ± 0.0003 | 4.5 ± 0.3 | 630 ± 30 | 1 | Star-forming galaxy | Pavesi et al. (2016) |
| J0129−0035 | 5.7787 ± 0.0001 | 1.99 ± 0.12 | 194 ± 12 | 1 | Quasar | Wang et al. (2013) |
| J1044−0125 | 5.7847 ± 0.0007 | 1.7 ± 0.3 | 420 ± 80 | 1 | Quasar | Wang et al. (2013) |
| J2310+1855 | 6.0031 ± 0.0002 | 8.83 ± 0.44 | 393 ± 21 | 1 | Quasar | Wang et al. (2013) |
| J0055+0146 | 6.0060 ± 0.0008 | 0.839 ± 0.132 | 359 ± 45 | 1 | Quasar | Willott et al. (2015a) |
| A383−5.1 | 6.0274 ± 0.0002 | 0.102 ± 0.032 | 100 ± 23 | 11.4 ± 1.9 | Lensed galaxy | Knudsen et al. (2016) |
| J2054−0005 | 6.0391 ± 0.0001 | 3.37 ± 0.12 | 243 ± 10 | 1 | Quasar | Wang et al. (2013) |
| J1319+0950 | 6.133 ± 0.0007 | 4.34 ± 0.6 | 515 ± 81 | 1 | Quasar | Wang et al. (2013) |
| VIMOS2911 | 6.1492 ± 0.0005 | 2.54 ± 0.13 | 264 ± 15 | 1 | Quasar | Willott et al. (2015a) |
| J2229+1457 | 6.1517 ± 0.0005 | 0.582 ± 0.075 | 351 ± 39 | 1 | Quasar | Willott et al. (2015a) |
| CLM1 | 6.1657 ± 0.0003 | 0.234 ± 0.031 | 162 ± 23 | 1 | Lyman–Break galaxy | Willott et al. (2015b) |
| SDSS J010013.02+280225.8 | 6.3258 ± 0.001 | 3.36 ± 0.46 | 300 ± 77 | 1 | Quasar | Wang et al. (2016) |
| HFLS3 | 6.3369 ± 0.0009 | 14.62 ± 3.05 | 470 ± 135 | 1 | Massive gas-rich galaxy | Riechers et al. (2013) |
| J2329−0301 | 6.4164 ± 0.0008 | 0.36 ± 0.04 | 477 ± 64 | 1 | Quasar | Willott et al. (2015a) |
| SDSS J114816.64+525150.3 | 6.42[a] | 3.9 ± 0.3 | 287 ± 28 | 1 | Quasar | Walter et al. (2009) |
| CFHQS J0210−0456 | 6.4323 ± 0.0005 | 0.269 ± 0.037 | 189 ± 18 | 1 | Quasar | Willott, Omont & Bergeron (2013) |
| PSOJ167−13 | 6.5157 ± 0.0008 | 3.84 ± 0.20 | 469 ± 24 | 1 | Quasar | Willott et al. (2017) |
| PSO J036.5078+03.0498 | 6.54122 ± 0.0018 | 5.2 ± 0.6 | 360 ± 50 | 1 | Quasar | Bañados et al. (2015) |
| P323+12 | 6.5881 ± 0.0003 | 1.05 ± 0.33 | 254 ± 48 | 1 | Quasar | Mazzucchelli et al. (2017) |
| J0305−3150 | 6.6145 ± 0.0001 | 3.44 ± 0.15 | 255 ± 12 | 1 | Quasar | Venemans et al. (2016) |
| P006+39 | 6.621 ± 0.002 | 0.78 ± 0.54 | 277 ± 161 | 1 | Quasar | Mazzucchelli et al. (2017) |
| P338+29 | 6.666 ± 0.004 | 1.72 ± 0.91 | 740 ± 541 | 1 | Quasar | Mazzucchelli et al. (2017) |
| RXJ1347:1216 | 6.7655 ± 0.0005 | 0.067 ± 0.012 | 75 ± 25 | 5.0 ± 0.3 | Lensed galaxy | Bradac et al. (2017) |
| J0109−3047 | 6.7909 ± 0.0004 | 2.04 ± 0.2 | 340 ± 36 | 1 | Quasar | Venemans et al. (2016) |
| COS-2987030247 | 6.8076 ± 0.0002 | 0.31 ± 0.04 | 124 ± 18 | 1 | Lyman–Break galaxy | Smit et al. (2018) |
| COS-3018555981 | 6.8540 ± 0.0003 | 0.39 ± 0.05 | 232 ± 30 | 1 | Lyman–Break galaxy | Smit et al. (2018) |
| J2348−3054 | 6.9018 ± 0.0007 | 1.57 ± 0.26 | 405 ± 69 | 1 | Quasar | Venemans et al. (2016) |
| ULAS J112001.48+064124.3 | 7.0842 ± 0.0004 | 1.03 ± 0.14 | 235 ± 35 | 1 | Quasar | Venemans et al. (2012) |
| J0313−1806 | 7.6423 ± 0.0013 | 0.6 ± 0.16 | 312 ± 94 | 1 | Quasar | Wang et al. (2021) |

*Note.*
[a]Owing to unavailable redshift uncertainty in the reference paper, we assign zero to the redshift uncertainty.

This paper has been typeset from a T$_E$X/L$^A$T$_E$X file prepared by the author.